# Redshift distortions of clustering: a Lagrangian approach

E. Hivon[1], F. R. Bouchet[1], S. Colombi[1,2], and R. Juszkiewicz[1,3,4]

[1] Institut d'Astrophysique de Paris, CNRS, 98 bis boulevard Arago, F-75014 Paris, France.
[2] NASA/Fermilab Astrophysics Center, Batavia, IL 60510, USA.
[3] Copernicus Center, Bartycka 18, PL-00716 Warsaw, Poland.
[4] Institute for Advanced Study, Olden Lane, Princeton, NJ 08540, USA.

**Abstract.** We study the effects of peculiar velocities on statistical measures of galaxy clustering. These effects occur when distances to the galaxies are estimated from their redshifts. It is assumed that the clustering pattern results from the gravitational instability of initially Gaussian, small-amplitude perturbations of a Friedman–Lemaître cosmological model. Explicit expressions are given for an arbitrary density parameter $\Omega$ of the model, both when the cosmological constant, $\Lambda$, is zero, and when the model is spatially flat, $\Omega + \Lambda/3H^2 = 1$.

Kaiser (1987) had analyzed the redshift distortion of the two–point correlation function. This function determines the variance of the density field distribution function and can be computed using linear perturbation theory. We show here how to compute higher order moments in redshift space, paying special attention to the skewness, or third moment of the density field, and its Fourier space counterpart, the bispectrum. This calls for a (weakly) *non–linear* analysis.

We rely on a perturbative expansion of particle trajectories in Lagrangian coordinates, using the formalism introduced by Moutarde et al. (1991) and further developed by Bouchet et al. (1992, 1994). This formalism extends to higher orders the Zel'dovich first order (i.e. linear) solution (1970). The lowest non-vanishing contribution to the skewness comes from the first and second-order terms in perturbation theory. Therefore, using Zel'dovich approximation would not be self-consistent and would yield inaccurate results. We show that a physically consistent and quantitatively accurate analysis of the growth skewness in redshift space can be obtained from second-order Lagrangian theory.

With practical applications to redshift surveys in mind, we also study the effects of spatial smoothing of the evolved density field. The necessary formalism was developed by Juszkiewicz and Bouchet (1991) and Juszkiewicz et al. (1993a). Here we give the first complete account of these calculations; we also extend the formalism by explicitly taking redshift distortions into account. We give analytic expressions for the gravitationally induced skewness as a function of the power spectrum and of $\Omega$, for a spherical top-hat and a Gaussian smoothing filter.

We compare our analytical predictions with measurements performed in numerical simulations, and find good agreement. These results should then prove useful in analyzing large scale redshift surveys. In particular, our results, in conjunction with the recent suggestion of Fry (1994), may solve a well known problem which always arises in conventional dynamical determinations of the mean density of the universe. Such studies produce estimates of $\Omega$ which are coupled with the parameters describing the bias in the galaxy distribution. As a result, a biased $\Omega = 1$ model is dynamically indistinguishable from an open, unbiased, one. For the first time, it may become possible to break this degeneracy, and decouple the estimates of linear and non-linear bias from the estimates of $\Omega$ and $\Lambda$.

## 1. Introduction

The appearance of structures is distorted in redshift space by peculiar velocities. At "small" scales, this leads to the "finger of god" effect: pattern are elongated along the line–of–sight due to their internal velocity dispersion. This is an intrinsically non–linear effect. At "large" scales, the effect is reversed: coherent inflows lead to a density contrast increase parallel to the line–of–sight. Indeed, foreground galaxies appear further than they are, while those in the back look closer, both being apparently nearer to the accreting structure.

Bean et al. (1983) and Davis and Peebles (1983) took advantage of the small scale distortion of clustering to estimate the pairwise velocity distortion. They were followed my many others. On the other hand, Sargent and Turner (1977) showed that the large scale distortion could be used to measure the cosmological parameter $\Omega$; Kaiser (1987) pointed out by using a

*Send offprint requests to*: E. Hivon

linear analysis that, in redshift space, each wave of wavevector $k$ and amplitude $\delta_k$ is amplified by a factor $(1 + f_1(\Omega)\mu_k^2)$, where $f_1 \sim \Omega^{0.6}$ describes the linear growth of perturbations [Peebles 1976, 1980 §14, and Eq. (16) below], and $\mu_k$ is the cosine of the angle between $k$ and the line–of–sight. The true power spectrum, $P(k) = \langle|\delta_k|^2\rangle$ in redshift space is replaced by

$$P_z(\boldsymbol{k}) = \left[1 + f_1(\Omega)\mu_k^2\right]^2 P(k); \qquad (1)$$

its Fourier transform, the two-point correlation function $\xi_2$, is strongly anisotropic in redshift space. This compression along the line–of–sight was computed by Lilje and Efstathiou (1989) and McGill (1990). Practical improvements to measure $\Omega$ were suggested by Hamilton (1992a), and used (Hamilton 1992b) to determine $\Omega$ from the IRAS 2 Jy redshift survey (Strauss et al. 1990, 1992). Recently, Fisher et al. (1994) measured the redshift space $\xi_2$ in the IRAS 1.2 Jy redshift survey. Many authors preferred though to measure the redshift space power spectrum $P_z(\boldsymbol{k})$ (see e.g. Feldman, Kaiser, and Peacock 1994; Cole, Fisher, and Weinberg 1994, and references therein).

Most of the analytical calculations performed in this field were done using a purely Eulerian approach. Still, Bouchet et al. (1992, hereafter BJCP), building on the work of Moutarde et al. (1991), developed a systematic perturbative approach from a Lagrangian point of view. This work was recently extended by Bouchet et al. (1994, hereafter BCHJ), complementing a parallel effort by Buchert (1994, and references therein). As we shall see, it is straightforward to use the Lagrangian approach to investigate the redshift distortion effect in the weakly non–linear regime. For example, Bouchet et al. (1992) indicated that the skewness factor $S_3$ (see below) of the unsmoothed density field is nearly invariant under the transformation to redshift space. The corresponding demonstration, which uses an "infinitely remote observer" approximation is given in Bouchet et al. (1994, hereafter BCHJ). In this paper we consider a slightly different approximation, valid in the limit of large samples, similar to the limit considered by Kaiser (1987). It gives a result, nearly identical to the one of BCHJ. We generalize the first order power spectrum results of Kaiser to second order quantities, such as the bispectrum $\langle\delta_{k_1}\delta_{k_2}\delta_{k_3}\rangle_{k_1+k_2+k_3=0}$ and its Fourier counterpart, $S_3$.

The skewness factor, $S_3$, is the ratio of the third moment of the probability distribution function (PDF) of the density contrast $\delta \equiv \rho/\langle\rho\rangle - 1$ to the square of its second moment (its variance)

$$S_3 \equiv \frac{\langle\delta^3\rangle}{\langle\delta^2\rangle^2}. \qquad (2)$$

Throughout this paper, the symbols $\langle\ldots\rangle$ denote ensemble averages, and the redshift space counterpart of a quantity $A$ is denoted by $A_z$. In (Eulerian) linear theory, a normal PDF remains normal, so that $S_3$ (as well as higher order factors $S_N$) is zero for an initially Gaussian field. Of course, linear theory (Eulerian or Lagrangian) is inadequate to describe indicators which are intrinsically of higher order. Under the influence of weakly–nonlinear gravitational instability, mode-mode couplings occur and establish a non-zero value for $S_3$, which remains constant until strongly non–linear effects come into play. For any smoothing scale, the ratio $S_3$ describes the asymmetry between the positive and negative tail of the PDF. As was shown recently by Juszkiewicz et al. (1993b), the low order moments of the PDF are enough to describe the weakly non–linear evolution of the PDF, thanks to an Edgeworth expansion in powers of the variance. The contribution of the third order moment i.e. of the skewness itself, can indeed already give a good idea of the overall shape of the PDF for $\langle\delta^2\rangle \lesssim 0.5$ (more precisely, the tails are reliably described within $\langle\delta^2\rangle^{-1/2}$ standard deviations from $\delta = 0$).

Juszkiewicz and Bouchet (1991, hereafter JB) and Juszkiewicz et al. (1993a, hereafter JBC) considered the effect of smoothing on $S_3$, when the *evolved* density contrast field $\delta$ is spatially averaged with a weight $W$, called "window function". In the unsmoothed limit obtained by Peebles (1980) for an $\Omega = 1$ model, $S_3$ depends neither on scale nor on the power spectrum, and equals to 34/7. But the skewness factor of the smoothed density field depends on both the shape of the window function and the local slope of the power spectrum. If $P(k)$ is not scale–invariant (as in CDM for instance), this induces a further scale dependence. For a top-hat window (a sphere with sharp boundaries) and a scale-invariant power spectrum with an index $n$ in the range $-3 \leq n < 1$, they obtained

$$S_3 = \frac{34}{7} + \frac{6}{7}(\Omega^{-2/63} - 1) - (n + 3) , \qquad (3)$$

and the above result was derived for real space. JBC considered Gaussian filters as well. Bernardeau (1993, 1994) generalized the result above for higher order $S_N$, $N \geq 4$, in the case of a top–hat smoothing filter, while the kurtosis in the Gaussian smoothing filter case was recently obtained by Lokas et al. (1994). These analytical results are fully confirmed by numerical simulations (e.g. Bouchet and Hernquist 1992, Weinberg and Cole 1992, JBC, Juszkiewicz et al. 1993b, Lucchin et al. 1994, Lokas et al. 1994). In this paper, we give the first complete account of the technique introduced by JB and JBC to estimate the skewness of the smoothed density field. These calculations, which here appear in the Appendix B, were earlier presented by JB and JBC in a rather terse form[1] When necessary, the formalism, presented here is also extended to combine smoothing effects with the effects of mapping from real space to redshift space. This enables comparisons between analytical predictions for an initially Gaussian field and practical measurements in the observed three-dimensional galaxy catalogs, such as those recently performed by Bouchet et al. (1991, 1993) in the IRAS 1.2Jy galaxy redshift catalog and by Gaztañaga (1992) in the CfA and SSRS redshift catalogues.

---

[1] Although we never used the the usual hypocritical "it is easy to show" phrase, our Letter to the Ap.J. (JBC) may have suggested that implicitly. Since it was not so easy at all (in fact it has taken several months to derive these results), and since we decided that it is awkward to keep distributing hand-written notes to colleagues, asking for details of our calculations, we decided to include them in this paper.

This paper is organized as follows. In Sect. 2, we recall the needed Lagrangian theory results and show how to derive the effect of peculiar velocities. We analyze the bispectrum in Sect. 3. Smoothing is considered in Sect. 4, where we compare our analytical results to numerical simulation measurements. Section 5 is the conclusion.

## 2. Redshift Distortion in Lagrangian Theory

We start by recalling some results of BJCP and BCHJ. The following section is devoted to the specific case of the redshift distortion.

### 2.1. Basics of the Perturbative Lagrangian Approach

We consider a non-relativistic pressureless fluid, in a Friedman-Lemaître background model in the the matter era. During that epoch, and at the scales of interest ($\lesssim$ 100Mpc), the perturbations can be treated as a ideal pressureless fluid in the Newtonian approximation (cf. Peebles, 1980). The scale factor of the metric is denoted by $a$. Throughout, we use comoving coordinates $x = r/a$, where $r$ stands for physical (expanding) coordinates. In the Eulerian approach, the primary quantity of the analysis is usually the density contrast field $\delta$. In the Lagrangian picture, each fluid element is tagged by its unperturbed Lagrangian (comoving) coordinate $q$, and the corresponding value of the displacement field $\Psi$. This vector field connects at any time the element's Eulerian and Lagrangian position through

$$x = q + \Psi(t, q). \tag{4}$$

The Jacobian of the transformation from $q$ to $x$ permits to express the requirement of mass conservation simply as $\rho(x)|\partial x/\partial q| d^3q = \rho(q)d^3q$, or

$$\delta[x(q)] = \left|\frac{\partial x}{\partial q}\right|^{-1} - 1. \tag{5}$$

In the weakly non–linear regime, solutions of the motion and field (Poisson) equations can be obtained by means of a perturbative expansion of the displacement field

$$\Psi = \varepsilon \Psi^{(1)} + \varepsilon^2 \Psi^{(2)} + \varepsilon^3 \Psi^{(3)} + \cdots, \tag{6}$$

where $\varepsilon$ is a formal parameter allowing to conveniently keep track of the various orders in the calculations. In the following, we need only the first two orders, which are separable:

$$\Psi^{(1)}(t, q) = g_1(t)\tilde{\Psi}^{(1)}(q), \tag{7}$$

$$\Psi^{(2)}(t, q) = g_2(t)\tilde{\Psi}^{(2)}(q). \tag{8}$$

The spatial part $\tilde{\Psi}^{(1)}(q)$ is the initial displacement field (this is Zel'dovich (1970) solution), while the second order obeys

$$\nabla_q \cdot \tilde{\Psi}^{(2)} = \sum_{i<j} \left[\frac{\partial \tilde{\Psi}_i^{(1)}}{\partial q_i}\frac{\partial \tilde{\Psi}_j^{(1)}}{\partial q_j} - \frac{\partial \tilde{\Psi}_i^{(1)}}{\partial q_j}\frac{\partial \tilde{\Psi}_j^{(1)}}{\partial q_i}\right], \tag{9}$$

with $\nabla_q = (\partial/\partial q_1, \partial/\partial q_2, \partial/\partial q_3)$. The assumption of irrotationnal flow leads to the constraints

$$\nabla_q \times \Psi^{(1)} = 0, \qquad \nabla_q \times \Psi^{(2)} = 0. \tag{10}$$

The displacement field $\Psi^{(1)}$ may be related to the initial Eulerian density contrast $\delta_i(x) = \mathcal{O}(\varepsilon)$ through the equation (5) expanded to linear order

$$\varepsilon \nabla_q \cdot \Psi^{(1)}(t, q) = \delta[x(q)] + \mathcal{O}(\varepsilon^2), \tag{11}$$

and evaluated at $t = t_i$ on the *l.h.s* and at $x = q$ on the *r.h.s.*.

The $n$-th order growth rates, $g_n$ depend on the matter content of the model. For a background model with $\Lambda = 0$, the fastest linear growing modes is identical to the well–known linear Eulerian growth rate $D(t)$ of the density contrast (e.g. Peebles 1980, BJCP), as can be seen from (11). The second order growth rate behaves according to

$$g_2 \simeq -\frac{3}{7}\Omega^{-2/63} g_1^2, \quad \Lambda = 0. \tag{12}$$

This approximation is better than 0.4% for $0.05 \lesssim \Omega \lesssim 3$ (BJCP). We will also consider a flat Universe with $\Lambda \neq 0$, that is $\Omega + \Lambda/3H^2 = 1$. The second order growth factor is then given by (BCHJ)

$$g_2 \simeq -\frac{3}{7}\Omega^{-1/143} g_1^2, \quad \Omega + \Lambda/3H^2 = 1. \tag{13}$$

Details of the derivations of these results may be found in BCHJ.

### 2.2. Redshift distortion in the Lagrangian framework

Let us now consider the case of spherical coordinates, when radial distances to the observer are estimated by means of redshift measurements. In that case, peculiar velocities distort the clustering pattern in redshift space, and the large scale infall increases the apparent density contrast. We will now use the Lagrangian formalism to derive an analytic expression, relating real space and redshift measurements, including weakly non–linear corrections.

The redshift space comoving position $z$ of a particle located in $r(q) = a\,x(q)$ is

$$z = \left(\frac{da}{dt}\right)^{-1} \left(\frac{dr}{dt} \cdot \frac{r}{r}\right) \frac{r}{r}. \tag{14}$$

Its modulus is then given at second order by

$$z = q_r + \varepsilon\,[1 + f_1(t)]\,g_1(t)\tilde{\Psi}_r^{(1)}(q) \\
+ \varepsilon^2\,[1 + f_2(t)]\,g_2(t)\tilde{\Psi}_r^{(2)}(q) + \mathcal{O}(\varepsilon^3), \tag{15}$$

where $q_r$ stands for $q \cdot r/r$. In the above expression, we have explicitly used the separability of $\Psi^{(1)}$ and $\Psi^{(2)}$, and defined the logarithmic derivatives of the growth rates

$$f_1 = (a/g_1)\,\partial g_1/\partial a, \quad \text{and} \quad f_2 = (a/g_2)\,\partial g_2/\partial a. \tag{16}$$

Limited expansions of $f_1$ and $f_2$ near $\Omega = 1$ yield, in the $\Lambda = 0$ case, $f_1 \asymp \Omega^{4/7}$ and $f_2 \asymp 2\Omega^{5/9}$. On the other hand, a better analytical approximation for $\Omega$ between 0.1 and 1 is given by

$$f_1 \approx \Omega^{3/5}, \quad \text{and} \quad f_2 \approx 2\,\Omega^{4/7}, \quad \text{when } \Lambda = 0. \quad (17)$$

The approximation for $f_1$ was proposed by Peebles (1976), while BJCP proposed the one for $f_2$. For a flat universe with $\Lambda \neq 0$, BCHJ suggest

$$f_1 \approx \Omega^{5/9}, \quad \text{and} \quad f_2 \approx 2\,\Omega^{6/11}, \quad \text{when } \Omega + \frac{\Lambda}{3H^2} = 1. \quad (18)$$

This $f_1$ term is very close to the fits previously proposed by Lahav et al. (1991) and Martel (1991).

With the coordinate transformation (15), we could directly compute the density contrast in redshift space according to the mass conservation requirement, that reads,

$$\delta_z(z(q)) = \left|\frac{\partial z}{\partial q}\right|^{-1} - 1. \quad (19)$$

Instead, we proceed in two steps: we map the Lagrangian $q$ coordinates onto the Eulerian ones $x$, and then proceed to map the latter onto our $z$ coordinates. We then see explicitly the terms brought by the real–to–redshift space mapping. Additionally, it makes it easier to handle the fact that the vector $q$ is in general not parallel to $r$, or in other words $q_r \neq q$. The mapping from real space to redshift space is then simplified by using the Eulerian coordinate $x = r/a$ as an intermediate step in the calculation, because $z \propto r$. After some algebra (see Appendix A.1), the density contrast in redshift space can be written as

$$\delta_z(z) = \varepsilon(\delta_1 + \Delta_1) + \varepsilon^2(\delta_2 + \Delta_2) + \mathcal{O}(\varepsilon^3), \quad (20)$$

with

$$\delta_1 = -\nabla_z \cdot \hat{\boldsymbol{\Psi}}^{(1)}, \quad (21)$$

$$\delta_2 = \left(\nabla_z \cdot \hat{\boldsymbol{\Psi}}^{(1)}\right)^2 + \hat{\boldsymbol{\Psi}}^{(1)} \cdot \nabla_z(\nabla_z \cdot \hat{\boldsymbol{\Psi}}^{(1)}) \quad (22)$$

$$- \left(1 + \frac{g_1^2}{g_2}\right) \nabla_z \cdot \hat{\boldsymbol{\Psi}}^{(2)}, \quad (23)$$

and

$$\Delta_1 = -f_1 \frac{\partial}{z^2 \partial z}(z^2 \hat{\Psi}_z^{(1)}) \quad (24)$$

$$\begin{aligned}
\Delta_2 &= f_1 \left[ \frac{\partial}{z^2 \partial z}(z^2 \hat{\Psi}_z^{(1)}) \nabla_z \cdot \hat{\boldsymbol{\Psi}}^{(1)} + \frac{\partial}{z^2 \partial z}(z^2 \hat{\boldsymbol{\Psi}}^{(1)} \cdot \nabla_z \hat{\Psi}_z^{(1)}) \right. \\
&\quad \left. + \hat{\Psi}_z^{(1)} \frac{\partial}{\partial z}(\nabla_z \cdot \hat{\boldsymbol{\Psi}}^{(1)}) \right] \\
&\quad + f_1^2 \left[ \frac{\partial}{z^2 \partial z}(z^2 \hat{\Psi}_z^{(1)}) \frac{\partial}{\partial z} \hat{\Psi}_z^{(1)} + 3\frac{[\hat{\Psi}_z^{(1)}]^2}{z^2} \right. \\
&\quad \left. + \hat{\Psi}_z^{(1)} \frac{\partial}{\partial z}\left\{\frac{\partial}{z^2 \partial z}(z^2 \hat{\Psi}_z^{(1)})\right\} \right] \\
&\quad - f_2 \frac{\partial}{z^2 \partial z}(z^2 \hat{\Psi}_z^{(2)}).
\end{aligned} \quad (25)$$

Note the introduction of the new displacement fields $\hat{\boldsymbol{\Psi}}^{(1)}$ and $\hat{\boldsymbol{\Psi}}^{(2)}$ defined by

$$\hat{\boldsymbol{\Psi}}^{(1)}(t,z) \equiv \boldsymbol{\Psi}^{(1)}(t,q=z), \quad \hat{\boldsymbol{\Psi}}^{(2)}(t,z) \equiv \boldsymbol{\Psi}^{(2)}(t,q=z). \quad (26)$$

By definition, they satisfy Eqs. (9), (10) and (11), but with all derivatives taken and calculations made in $z$ space instead of $q$ space. In real space, i.e., when one is interested in the calculation of $\delta(x)$, the terms $\Delta_1$ and $\Delta_2$ vanish, which is equivalent to setting $f_1 = f_2 = 0$, and the displacements $\hat{\boldsymbol{\Psi}}^{(1)}(t,z)$ and $\hat{\boldsymbol{\Psi}}^{(2)}(t,z)$ must be replaced by the Eulerian displacement fields $\bar{\boldsymbol{\Psi}}^{(1)}(t,x) \equiv \boldsymbol{\Psi}^{(1)}(t,q=x)$ and $\bar{\boldsymbol{\Psi}}^{(2)}(t,x) \equiv \boldsymbol{\Psi}^{(2)}(t,q=x)$.

The expressions $\Delta_1$ and $\Delta_2$ contain terms of the form $F(z)/z$ or $F(z)/z^2$, where $F(z)$ is a quantity without preferred symmetry nor direction, because we assume the universe is spatially homogeneous and isotropic. When $z$ is large compared to the scale $\ell$ considered (which can be for instance a smoothing scale, a separation between two objects, or simply a Fourier space wavelength), these terms can be neglected compared to those in $F(z)$. From a statistical point of view, this requires the volume sampled, $V$, to have a size much larger than $\ell$. In that case, when statistical quantities are measured in that volume $V$, even if the small $z$ contributions are taken into account, the terms in $\langle F(z)/z \rangle$ and $\langle F(z)/z^2 \rangle$ have a negligible weight. We now assume this "large volume" limit. The quantities $\Delta_1$ and $\Delta_2$ then simplify dramatically and read

$$\Delta_1 = -f_1 \frac{\partial}{\partial z} \Psi_z^{(1)}, \quad (27)$$

$$\begin{aligned}
\Delta_2 &= f_1 \frac{\partial}{\partial z}\{\nabla_z \cdot (\hat{\Psi}_z^{(1)} \hat{\boldsymbol{\Psi}}^{(1)})\} + f_1^2 \frac{\partial}{\partial z}\left\{\hat{\Psi}_z^{(1)} \frac{\partial \Psi_z^{(1)}}{\partial z}\right\} \\
&\quad - f_2 \frac{\partial}{\partial z} \hat{\Psi}_z^{(2)}.
\end{aligned} \quad (28)$$

Our assumption is *almost* equivalent to the "infinitely remote observer" approximation used by BCHJ. But there is a significant difference: instead of using a Cartesian grid deformed along the line-of-sight, we keep the radial feature of the redshift distortion, i.e., we remain in spherical coordinates.

## 3. Redshift distortion of the bispectrum

In this section, we use a statistical approach and work in Fourier space. We first compute the bispectrum $B(k_1, k_2, k_3) \equiv \langle \delta_{k_1} \delta_{k_2} \delta_{k_3} \rangle_{k_1 + k_2 + k_3 = 0}$ in redshift space, in the weakly non–linear regime. Then we quantify the effect of the redshift distortion by using the reduced amplitude $Q$ of the bispectrum which was introduced by Fry & Seldner in 1982 (hereafter FS).

The initial fluctuations are assumed to be Gaussian-distributed. In Fourier space, this is equivalent to

$$\langle \hat{\delta}_i(k)\hat{\delta}_i(l) \rangle = \delta_{\text{Dirac}}(k+l) P_i(k), \quad (29)$$

$$\langle \hat{\delta}_i(k)\hat{\delta}_i(l)\hat{\delta}_i(m) \rangle = 0, \quad (30)$$

$$\langle \hat{\delta}_i(k)\hat{\delta}_i(l)\hat{\delta}_i(m)\hat{\delta}_i(n) \rangle = \delta_{\text{Dirac}}(k+l)\,\delta_{\text{Dirac}}(m+n) \\
P_i(k)P_i(m) + \text{perm.} \quad (31)$$

and so on. Here "perm." stands for the terms obtained by pair permutations of $k, l, m, n$ (3 terms in all). The function $\hat{\delta}_i(k)$ is the Fourier transform of the initial density contrast $\delta_i(x)$, and $P_i(k) \equiv <|\hat{\delta}_i(k)|^2>$ is its power spectrum.

It is convenient to evaluate the quantities $\delta_1, \delta_2, \Delta_1$ and $\Delta_2$ defined in Sect. 2.2 by using Fourier analysis. The corresponding integrals are given in Appendix A.2 [Eqs. (A16), (A18), (A19), (A20)]. After taking their ensemble averages, we get at lowest order in $\varepsilon$,

$$P_z(k) = b_1 P(k), \quad (32)$$

with

$$b_1 = 1 + \frac{2}{3}f_1 + \frac{1}{5}f_1^2. \quad (33)$$

This is nothing else but the result of Kaiser [1977, Eq. (1)], when averaged over angles. The bispectrum in redshift space is given by

$$\begin{aligned}
B_z(\mathbf{k}_1, \mathbf{k}_2, \mathbf{k}_3) &= P(k_1)P(k_2) \Big[ a_0 + a_2 \cos^2 \theta_{12} + a_4 \cos^4 \theta_{12} \\
&\quad + (a_1 \cos \theta_{12} + a_3 \cos^3 \theta_{12}) \left( \frac{k_1}{k_2} + \frac{k_2}{k_1} \right) \\
&\quad + (1 - \cos^2 \theta_{12})^2 \left( \frac{x_1 (k_1^2 + k_2^2) + x_2 k_1 k_2 \cos \theta_{12}}{k_1^2 + k_2^2 + 2 k_1 k_2 \cos \theta_{12}} \right) \Big] \\
&\quad + \text{cyc.} \quad (34)
\end{aligned}$$

where $\cos \theta_{ij} \equiv \mathbf{k}_i . \mathbf{k}_j / (k_i k_j)$. Here "cyc." stands for the terms obtained by cyclic permutation of the triple-valued indices. The factors $a_i$, $i = 0, \ldots, 4$, $x_1$ and $x_2$ depend on the values of $\Omega$ and $\Lambda$. We list them here:

$$\begin{aligned}
a_0 &= (1 - \kappa) + \frac{2}{3}(2 - \kappa)f_1 + \frac{1}{15}(11 - \kappa)f_1^2 + \frac{6}{35}f_1^3 + \frac{6}{315}f_1^4 \\
&\quad - \kappa f_2 (\frac{1}{3} + \frac{6}{15}f_1 + \frac{1}{35}f_1^2), \\
a_1 &= 1 + \frac{4}{3}f_1 + \frac{14}{15}f_1^2 + \frac{12}{35}f_1^3 + \frac{15}{315}f_1^4, \\
a_2 &= (1 + \kappa) + \frac{2}{3}(2 + \kappa)f_1 + \frac{1}{15}(23 - \kappa)f_1^2 + \frac{30}{35}f_1^3 + \frac{48}{315}f_1^4 \\
&\quad + \kappa f_2 (\frac{1}{3} + \frac{6}{15}f_1 - \frac{3}{35}f_1^2), \\
a_3 &= \frac{4}{15}f_1^2 + \frac{8}{35}f_1^3 + \frac{20}{315}f_1^4, \\
a_4 &= \frac{2}{15}(1 + \kappa)f_1^2 + \frac{4}{35}f_1^3 + \frac{16}{315}f_1^4 + \frac{4}{35}\kappa f_1^2 f_2, \\
x_1 &= \frac{2}{15}\kappa f_1 f_2, \\
x_2 &= -\frac{4}{35}\kappa f_1^2 f_2, \quad (35)
\end{aligned}$$

and

$$\kappa = g_2/g_1^2. \quad (36)$$

Again, we can set $f_1 = f_2 = 0$ to have the value of the bispectrum in real space and get the well known expression (obtained by Fry in 1984 for $\Omega = 1$)

$$\begin{aligned}
B(k_1, k_2, k_3) &= P(k_1)P(k_2) \Big[ 2 + \cos \theta_{12} \left( \frac{k_1}{k_2} + \frac{k_2}{k_1} \right) \\
&\quad - \left( 1 + \frac{g_2}{g_1^2} \right) (1 - \cos^2 \theta_{12}) \Big] + \text{cyc.} \quad (37)
\end{aligned}$$

The values of $a_i/b_1^2$, $x_j/b_1^2$ are listed in Table 1 for $(\Omega, \Lambda/3H^2) = (1, 0), (0.1, 0)$ and $(0.1, 0.9)$. Except when $\Omega = 1$, we have $(a_i/b_1^2)_z \simeq (a_i/b_1^2)_{\text{real}}$. In other words, when $\Omega$ is sufficiently small compared to unity, and $\Lambda$ is arbitrary, we obtain

$$B_z(k_1, k_2, k_3) \simeq b_1^2 B(k_1, k_2, k_3). \quad (38)$$

When $\Omega$ approaches unity, the differences between real space and redshift space are mainly introduced by the factors $a_0/b_1^2$, $a_2/b_1^2$ and $a_3/b_1^2$.

**Table 1.** Factors $b_1$, $a_i$ and $x_j$ involved in the computation of the power-spectrum and the bispectrum for various values of $\Omega$

| Fact. | redshift space | | | real space | | |
|---|---|---|---|---|---|---|
| | $\Omega = 1$ | $\Omega = .1$ | $\Omega = .1$[a] | $\Omega = 1$ | $\Omega = .1$ | $\Omega = .1$[a] |
| $b_1$ | 1.867 | 1.180 | 1.201 | 1 | 1 | 1 |
| $a_0/b_1^2$ | 1.335 | 1.459 | 1.429 | 1.429 | 1.461 | 1.436 |
| $a_1/b_1^2$ | 1.050 | 1.005 | 1.006 | 1 | 1 | 1 |
| $a_2/b_1^2$ | 1.043 | .5770 | .6145 | .5714 | .5389 | .5644 |
| $a_3/b_1^2$ | .1603 | .0149 | .0180 | 0 | 0 | 0 |
| $a_4/b_1^2$ | .0411 | .0034 | .0044 | 0 | 0 | 0 |
| $-x_1/b_1^2$ | .0328 | .0060 | .0064 | 0 | 0 | 0 |
| $x_2/b_1^2$ | .0281 | .0013 | .0015 | 0 | 0 | 0 |

[a] For a flat Universe with $\Lambda/3H^2 = .9$

To illustrate our results, it is convenient to use the reduced amplitude $Q$ of the bispectrum. It is defined by (FS)

$$Q(\mathbf{k}_1, \mathbf{k}_2, \mathbf{k}_3) \equiv \frac{B(\mathbf{k}_1, \mathbf{k}_2, \mathbf{k}_3)}{P(k_1)P(k_2) + P(k_2)P(k_3) + P(k_3)P(k_1)}. \quad (39)$$

The parameter $Q$, defined above, must satisfy Eq. (34) and the constraint $\mathbf{k}_1 + \mathbf{k}_2 + \mathbf{k}_3 = 0$. Therefore, $Q$ can be described as a function of $k_1, k_2$ and $\cos \theta \equiv \cos \theta_{12}$. It also depends on the shape of the initial power spectrum and was first measured in the galaxy distribution by FS. Comparisons of theoretical predictions with $N$-body experiments were done by Fry et al. (1993). In the weakly non–linear regime, they found a good agreement between the numerical measurements and the predictions of second order perturbation theory.

Figure 1 gives $Q$ as a function of $\theta$ for $k_1/k_2 = 2$ and various values of $(\Omega, \Lambda)$ [$(1, 0), (0.1, 0)$ and $(0.1, 0.9)$]. We have considered 5 scale–invariant power-spectra, with $n = -3, -2$,

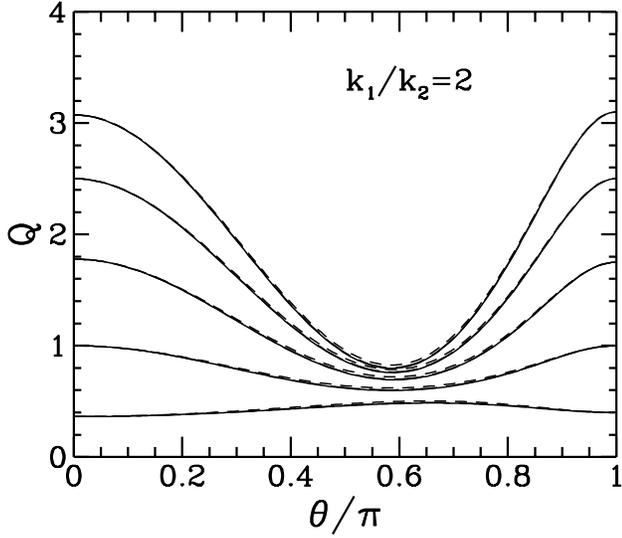

**Fig. 1.** The reduced amplitude $Q$ of the bispectrum in real space versus the wavevectors angle $\theta$ in the case $k_1/k_2 = 2$, for $\Omega = 1$ (solid line), $\Omega = 0.1$ and $\Lambda = 0$ (short dashes) and $\Omega = 0.1$ and $\Omega + \Lambda/3H^2 = 1$ (long dashes). Scale–Invariant power spectra are considered with $n = -3$, -2, -1, 0, +1 (from top to bottom).

-1, 0 and +1. In that case, $Q \equiv Q(k_1/k_2, \cos\theta_{12}, n, \Omega)$. As expected, the dependence of $Q$ on $\Omega$ and $\Lambda$ is very weak. The first panel of Fig. 2 is the redshift space analog of Fig.1. We have kept the solid curves, which give $Q$ in real space for $\Omega = 1$. As we already saw above, there is a significant difference between the real and redshift space values of $Q$ only when $\Omega \gtrsim 1$. This difference can be as large as $\sim 20\%$, but the general features of $Q$ versus $\theta$ and $k_1/k_2$ are not changed, as confirmed by the second and third panels of Fig. 2 which shows $Q$ versus $\theta$ for $k_1/k_2 = 1$ and for $k_1/k_2 = 10$ in real and redshift space when $\Omega = 1$.

Finally, when $k_1/k_2 = 2$, the ratio $Q_z/Q$ depends only weakly on the initial power spectrum, as can be seen in Fig. 3. We find that the $\Omega$–dependence of this ratio for $\Omega \lesssim 1$ is given by

$$Q_z(k_1, k_2, k_3) = Q(k_1, k_2, k_3)[1 + \Delta(k_1, k_2, k_3)], \quad (40)$$

where $\Delta(k_1, k_2, k_3)$ is a smoothly varying function such that

$$|\Delta(k_1, k_2, k_3)| \lesssim 0.2\,\Omega. \quad (41)$$

So far, all of the quantities derived here were statistical measures of the spatial distribution of matter. However, if the galaxy distribution is biased with respect to that of matter, the reduced moments (like the skewness), estimated from counts of galaxies do not have to coincide with those for the matter. Following Fry (1994), we will now rederive the effect of a *local biasing* on the $Q(k_1/k_2, \theta, \Omega, n)$ parameter. The local biasing prescription assumes that the number–density field of galaxies, $n_g(\boldsymbol{x})$ is a smooth local function $f$ of the underlying density field of the

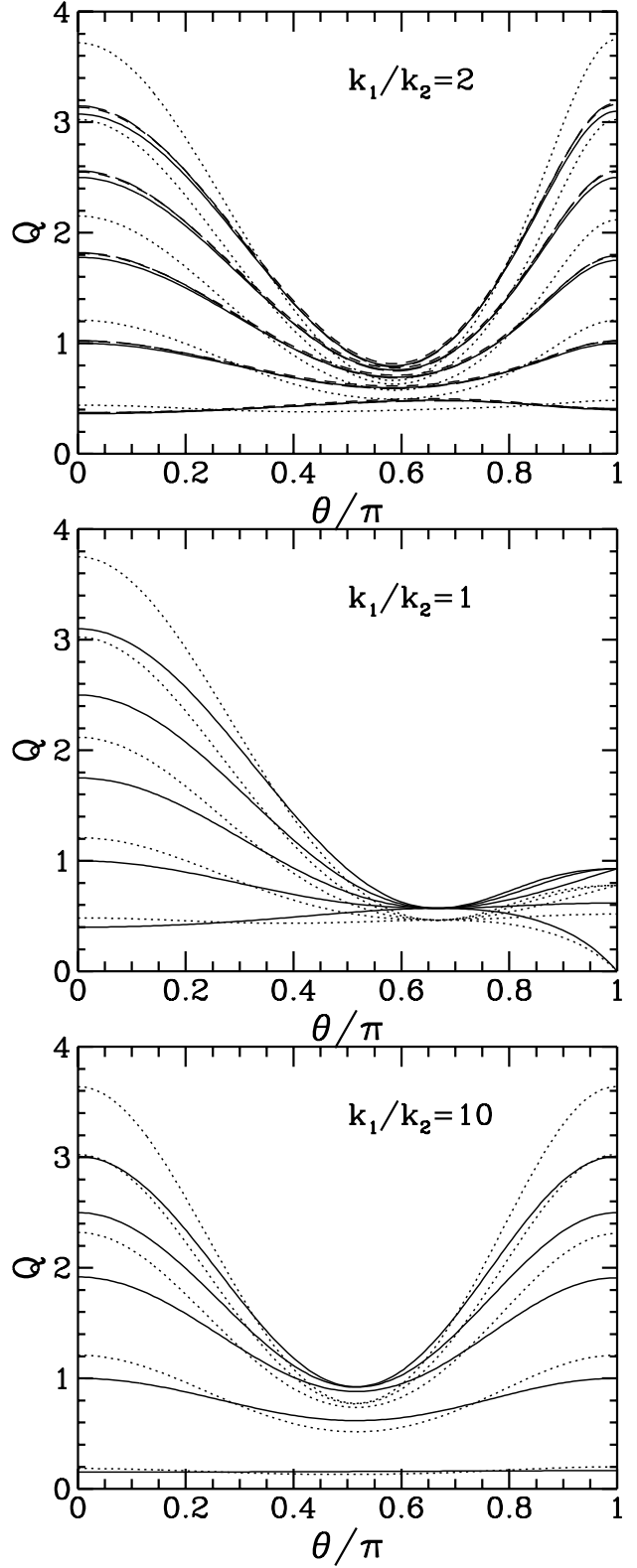

**Fig. 2.** The reduced amplitude $Q$ of the bispectrum as a function of the wavevectors angle $\theta$, in real and redshift space for several values of $k_1/k_2$. As in Fig. 1, we consider scale–invariant power-spectra with $n = -3$ (top curve), -2, -1, 0, +1 (bottom curve). In all panels, the full line and the dotted line refer respectively to real and redshift space for $\Omega = 1$. In the top panel, the short dashes correspond to $\Omega = 0.1$, $\Lambda = 0$ and the long dashes to $\Omega = 0.1$, $\Omega + \Lambda/3H^2 = 1$, in redshift space.

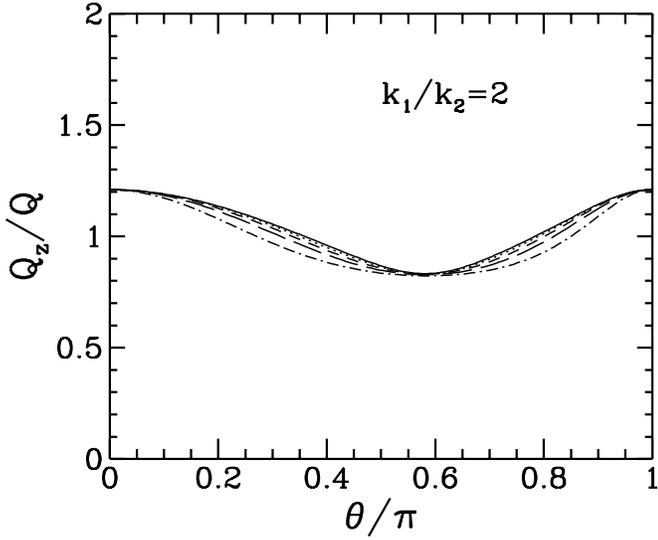

**Fig. 3.** The ratio $Q_z/Q$ as a function of angle for $k_1/k_2 = 2$, $\Omega = 1$ and several scale–invariant power-spectra, with $n = -3$ (full line), $n = -2$ (dots), $n = -1$ (short dashes), $n = 0$ (long dashes), $n = +1$ (dot-dashes).

matter, $\rho(\mathbf{x})$, that is $n_g(\mathbf{x}) = f[\rho(\mathbf{x})]$. This allows a Taylor expansion

$$\delta_g = b\delta + \Sigma_{m=2}^{\infty} \frac{b_m \delta^m}{m!} \tag{42}$$

where $b$ is the usual "linear bias", while higher order terms allow for nonlinear biasing. They are also necessary to ensure that each term in the perturbative expansion for $\delta_g$ has a zero mean value (cf. Fry & Gaztañaga 1993; Juszkiewicz et al. 1993b). To derive the relationship between $Q$ and its estimate based on galaxy counts, $Q_g$, it is enough to substitute the expansion for $\delta_g(\delta)$ into the perturbative series for $\delta$, then rederive the bispectrum and the $Q$ parameter. The result of all this is simply given by (Fry 1994)

$$Q_g = \frac{Q}{b} + \frac{b_2}{b^2}. \tag{43}$$

Thus both the linear bias $b$ and the non-linear bias $b_2$ can be determined by comparing the measured shape of $Q_g(k_1/k_2, \theta)$ with the theoretical predictions for the underlying matter field (assuming that indeed large scale structures arose from the gravitational development of Gaussian initial conditions, which can be tested otherwise). This may solve a well known problem which always arises in conventional dynamical determinations of the mean density of the universe. Such studies tend to produce only estimates of $\Omega^{0.6}/b$, where the linear bias parameter, $b$, is unknown (see, e.g. Strauss et al. 1992). As a result, a biased $\Omega = 1$ model is dynamically indistinguishable from an open unbiased one. Therefore, in conjunction with our calculations, which allow to properly take into account the effects of redshift distortion on $Q$, Fry's result may for the first time provide a way to use redshift surveys to break the degeneracy, and decouple the estimates of linear and non-linear bias from the estimates of $\Omega$ and $\Lambda$.

## 4. Skewness of the smoothed density field in redshift space

We now leave Fourier space and turn to the skewness factor $S_3 \equiv \langle \delta^3 \rangle / \langle \delta^2 \rangle^2$ of the smoothed density field. The smoothing windows considered here are the most widely used ones, i.e. the Gaussian window and the top-hat window. Firstly, we study analytically the redshift space distortion in the weakly non–linear regime. Then, we compare analytical results with careful measurements in $N$–body experiments and estimate a dynamical range of validity of these results.

### 4.1. Analytical predictions in the weakly non–linear regime

In the weakly non–linear regime, one can use second order perturbation theory to compute $S_3$ (JB, JBC and Appendix A.2). By using Fourier integrals, the low orders moments of the density field $\langle \delta^2 \rangle$ and $\langle \delta^3 \rangle$ can be written in terms of, respectively, the power spectrum and the bispectrum. This gives

$$\sigma^2(\ell) \equiv \langle \delta^2 \rangle = \int \frac{d\mathbf{k}}{(2\pi)^3} P(k) W_\ell^2(k), \tag{44}$$

$$\langle \delta^3 \rangle = \iint \frac{d\mathbf{k}_1 d\mathbf{k}_2}{(2\pi)^6} B(\mathbf{k}_1, \mathbf{k}_2, -\mathbf{k}_1 - \mathbf{k}_2) \\ W_\ell(k_1) W_\ell(k_2) W_\ell(|\mathbf{k}_1 + \mathbf{k}_2|), \tag{45}$$

where $W_\ell(k)$ stands for the Fourier transform of the smoothing filter of characteristic size $\ell$. These expressions are of course also valid in redshift space. But then the quantities $P(k)$ and $B(\mathbf{k}_1, \mathbf{k}_2, \mathbf{k}_3)$ have to be replaced by their redshift space counterparts $P_z(k)$ and $B_z(\mathbf{k}_1, \mathbf{k}_2, \mathbf{k}_3)$. Appendix B gives a full length account of the computation of the redshift space skewness (and hence of the real space if one takes $f_1 = f_2 = 0$) for scale–invariant initial power-spectra, with or without smoothing. Here we just give the main results. We first consider the case without smoothing. In some cases, the integrals in Eqs. (44), (45) may diverge. In order to insure convergence, one then needs to introduce a large–$k$ cutoff in the power spectrum. Nevertheless, the ratio $S_3 = \langle \delta^3 \rangle / \langle \delta^2 \rangle^2$ may still be well defined, provided it remains finite when the cutoff tends to infinity. In redshift space, it is given by

$$S_{3,z} = \frac{3}{b_1^2}\left(a_0 + \frac{a_2}{3} + \frac{a_4}{5} + \tilde{X}\right), \tag{46}$$

where $\tilde{X}$ depends on the shape of the power spectrum, and is hard to compute analytically. Nevertheless, it can usefully be bounded as follows

$$\frac{29}{20}x_1 - \frac{23}{120}x_2 < \tilde{X} < \frac{19}{20}x_1 + \frac{7}{120}x_2. \tag{47}$$

By taking $f_1 = f_2 = 0$ in Eq. (46), (35), and (36) we recover the familiar value of $S_3$ in real space, that is

$$S_3 = 4 - 2\frac{g_2}{g_1^2}, \tag{48}$$

which is independent of the initial power spectrum and nearly independent of $\Omega$, as found by BJCP and JBC. The following values of $S_{3,z}$,

$$\Omega = 1, \Lambda = 0, \quad S_{3,z} = 5.01 \pm 0.02, S_3 = 4.86,$$
$$\Omega = .1, \Lambda = 0, \quad S_{3,z} = 4.95 \pm 0.01, S_3 = 4.92, \quad (49)$$
$$\Omega = .1, \Lambda/3H^2 = .9, \quad S_{3,z} = 4.89 \pm 0.01, S_3 = 4.87,$$

show that the projection in redshift space does not significantly change the value of the skewness factor of the unsmoothed density field. This confirms the results of BJCP and BCHJ. Indeed, by using their "infinitely remote observer" approximation, BJCP and BCHJ had found $S_{3,z} = 5.07 \pm 0.07$ for $\Omega = 1$ and $S_{3,z} = 4.96 \pm 0.03$, for $\Omega = 0.1$ ($\Lambda = 0$). These results are in a good agreement with those obtained here, as expected.

Numerical values of $S_3$ in real and redshift space for a Gaussian and a top-hat smoothing are listed in Tables 2 and 3. Figure 4 summarizes the behavior of skewness when the spectral index $n$ is varied, for background model with $\Lambda = 0$, and $\Omega = 0.1$ or $\Omega = 1$. JB and JBC discussed rather extensively the $n$ dependence of $S_3$, and JBC had found that $S_3$ is nearly independent of $\Omega$. Here we note that this remains true for spatially-flat universes with non-vanishing cosmological constant.

The mapping from real to redshift space brings only very small changes, except for values of $\Omega$ close to unity. In the case $\Omega = 1$, the difference between $S_{3,z}$ and $S_3$ increases with $n$, but remains small, at most of order 15%. This behavior is maybe slightly different from what one could have naïvely expected from the analysis of the bispectrum (e.g. Fig. 3). In that case, indeed, the effect of a redshift space projection appeared to be of the same order for any value of $n$. For a top-hat filter, the linear relation between $S_3$ and $n$ in real space maps into another linear relation between $S_{3,z}$ and $n$. For example, when $\Omega = 1$, $S_{3,z}$ is well fitted by

$$S_{3,z} \simeq \frac{35.2}{7} - 1.15(n+3). \tag{50}$$

This is to be compared with the real space values given by Eq. (3).

**Table 2.** Skewness in redshift and real space for different power spectra $P(k) = k^n$ with Gaussian smoothing

| | Skewness | | | | | |
|---|---|---|---|---|---|---|
| | in redshift space | | | in real space | | |
| $n$ | $\Omega = 1$ | $\Omega = .1$ | $\Omega = .1$[a] | $\Omega = 1$ | $\Omega = .1$ | $\Omega = .1$[a] |
| −3 | 5.022 | 4.947 | 4.894 | 4.857 | 4.922 | 4.871 |
| −2 | 4.043 | 4.1 | 4.042 | 4.022 | 4.089 | 4.036 |
| −1 | 3.364 | 3.54 | 3.475 | 3.468 | 3.541 | 3.484 |
| 0 | 2.914 | 3.217 | 3.139 | 3.144 | 3.228 | 3.162 |
| 1 | 2.652 | 3.108 | 3.012 | 3.029 | 3.132 | 3.051 |

[a] For a flat Universe with $\Lambda/3H^2 = .9$

**Table 3.** Skewness in redshift and real space for different power spectra $P(k) = k^n$ with top-hat smoothing

| | Skewness | | | | | |
|---|---|---|---|---|---|---|
| | in redshift space | | | in real space | | |
| $n$ | $\Omega = 1$ | $\Omega = .1$ | $\Omega = .1$[a] | $\Omega = 1$ | $\Omega = .1$ | $\Omega = .1$[a] |
| −3 | 5.022 | 4.947 | 4.894 | 4.857 | 4.922 | 4.871 |
| −5/2 | 4.443 | 4.439 | 4.384 | 4.357 | 4.422 | 4.371 |
| −2 | 3.868 | 3.932 | 3.876 | 3.857 | 3.922 | 3.871 |
| −3/2 | 3.294 | 3.425 | 3.367 | 3.357 | 3.422 | 3.371 |
| −1 | 2.72 | 2.917 | 2.858 | 2.857 | 2.922 | 2.871 |
| −1/2 | 2.144 | 2.41 | 2.349 | 2.357 | 2.422 | 2.371 |
| 0 | 1.557 | 1.902 | 1.839 | 1.857 | 1.922 | 1.871 |

[a] For a flat Universe with $\Lambda/3H^2 = .9$

### 4.2. Measurements of $S_3$ in $\Omega = 1$, scale–invariant, N–body simulations

Numerous previous comparisons with $N$–body simulations showed a very good agreement with second order perturbation theory predictions up to a (density contrast) variance of order unity (see, e.g., JBC and Juszkiewicz et al. 1993b). However, all these measurements were made in real space. Here we extend these tests to the predictions concerning the redshift distortion in order to determine their range of validity. In particular, it is important to find out, when strongly non–linear effects, like the "fingers of god", become large enough to invalidate our perturbative results. Section 4.2.1 describes the simulations done, the measurements method, and our error estimates, while section 4.2.2 discusses the results.

#### 4.2.1. Measurement: method and uncertainties

Using the PM code of Moutarde et al. (1991), we performed three $N$–body simulations with $\Omega = 1$, $64^3$ matter particles in a $128^3$ grid, and scale–invariant initial conditions with $P(k) = A\,k^n$, $n = 0, -1, -2$. Following Efstathiou et al. (1988), we picked a normalization of the initial power spectrum so that it matched the white-noise level at the Nyquist frequency of the particle grid, except for $n = -2$. In that case, the normalization constant $A$ was chosen 4 times smaller. The density field was estimated by convolving the particle distribution with either a Gaussian filter of half-width $\ell$, or a spherical window of radius $\ell$ (all lengths are expressed in units of the size of the simulation box $L_{\text{box}} \equiv 1$).

In principle, the only relevant scale in the simulations is the correlation length $\ell_0$, for which the two-body correlation function $\xi_2$ is unity. Thus measurements made at different times of a simulation should give statistically equivalent results provided there are made at the same fixed fraction of that (time–dependent) scale $\ell_0$. As far as $S_3$ is concerned, it should thus be constant at constant variance (for a given value of $n$). If the expected scaling behavior is only approximate, it may come from one or several of the following numerical limitations:

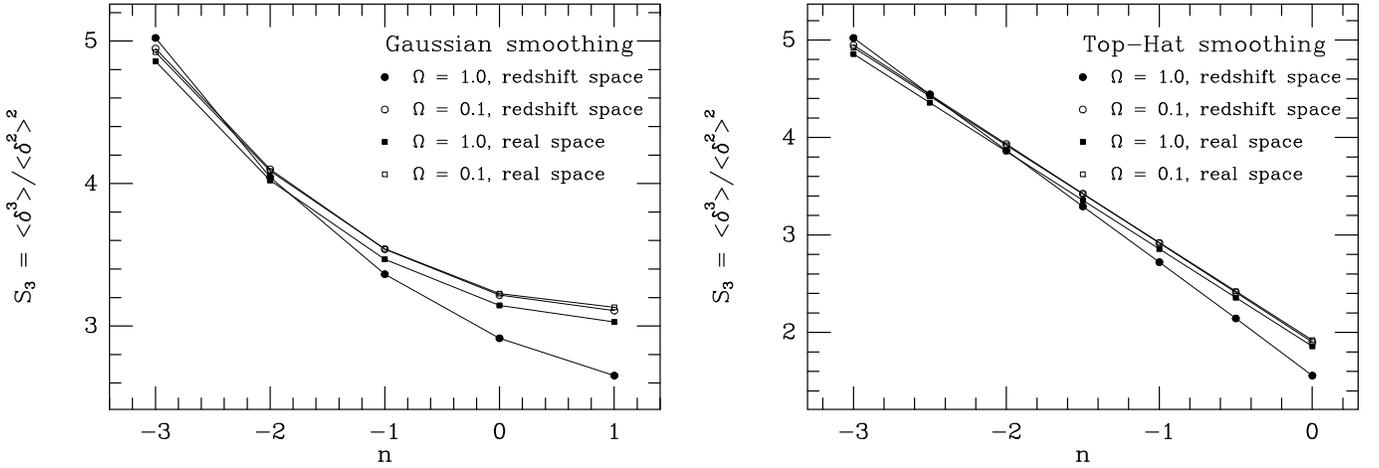

**Fig. 4.** Skewness factor in real and redshift space, for Gaussian and top-hat smoothing as a function of the power spectrum index $n$.

1. *Initial conditions:* they were created by slightly displacing matter particles from a regular pattern by using the Zel'dovich approximation. This is not totally innocuous. Firstly, the constraint arising from this regular pattern may contaminate the latter measurements (in particular in underdense regions, see e.g. Bouchet et al. 1991, Bouchet & Hernquist 1992, Colombi et al. 1994). Secondly, the Zel'dovich prescription is only valid at first (linear) order, and thus does not set up properly second order quantities such as the skewness (JBC). One thus has to wait long enough for transients to become negligible. Typically, the correlation length needs to be larger than the mean interparticle distance. This means that we should have

$$\ell_0 \gtrsim 1/64 \simeq 0.0156. \tag{51}$$

2. *Finite spatial resolution:* Bouchet et al. (1985) showed that, in the linear regime, the effect of the grid could affect the linear growth of fluctuations on scales substantially larger than the grid scale $\Delta x$ (say $8\Delta x$ for a typical PM code without a staggered mesh, if better than 10% accuracy per mode is required). On the other hand, the transfer of power in the gravitational clustering problem appears to proceed from the large to the small scale. This means that "gravity is forgiving", and reasonably accurate results in the strongly non–linear regime can be obtained even on scales comparable to the cell size (e.g. Weinberg et al. 2001, but PM velocities may systematically be underestimated by as much as 20%). Thus provided that our measurements are such that the correlation length $\ell_0$ remains large as compared to the grid scale (but small as compared to the box size), our skewness measurements should be accurate. Indeed, this insures that measurements on scales comparable to the grid scale ($\log_{10} \ell \gtrsim -2.1$) are in the strongly non–linear regime, and the weakly non–linear measurements are on much larger scales, where a PM code is reasonably accurate. Furthermore, we compared our results to those obtained from high resolution simulations (made with the tree code of Bouchet & Hernquist, 1988) with identical initial conditions (Colombi et al. 1994). The differences are rather small, of order at most ten percent; the measured $S_3$ in the high resolution simulations are only slightly smaller than the measured $S_3$ in our PM simulations.

3. *Finite volume effects:* since we used periodic boundary conditions, the power coming from scales larger than the box size is missing. Also, at scales smaller but close to the box size, only a few independent modes of the power spectrum are sampled. This increases the uncertainties on the measurements as one approaches the box size. To minimize these effects, one usually imposes $\log_{10} \ell \lesssim -1$. But this constraint is not necessarily sufficient. Indeed, the large $\rho$ tail of the PDF is determined by just a few large clusters and that tail is thus subject to fluctuations due to small number statistics, till it reaches an artificial cutoff at $\rho_{\max}$ (Colombi et al. 1994, hereafter CBS). If the volume is large enough, these modifications of the true tail happen for such small values of the PDF that is has little effect (however, higher order statistics are more sensitive to this scale-dependent effect than those of lower order). If the correlation length, which gives the typical size of a cluster, is small compared to the box size, then this effect should also be small and can be corrected for (or at least the corresponding error–bar can be evaluated, see CBS and Appendix C).

4. *Discreteness effects:* starting from a discrete particle representation, we want to measure statistical properties of the underlying continuous density field. This shot noise can easily be corrected for a top–hat filter and is naturally erased by a Gaussian filter. It can be neglected at scales large compared to the typical distance between two particles in a cluster (e.g., Balian & Schaeffer 1989).

The considerations above lead us to measure $\sigma^2$ and $S_3$ in the following intervals

$$-2.1 \leq \log_{10} \ell \leq -1.3, \quad \Delta \log_{10} \ell = 0.2, \tag{52}$$

for a Gaussian filter and

$$-2.0 \leq \log_{10} \ell \leq -1.2, \quad \Delta \log_{10} \ell = 0.2, \tag{53}$$

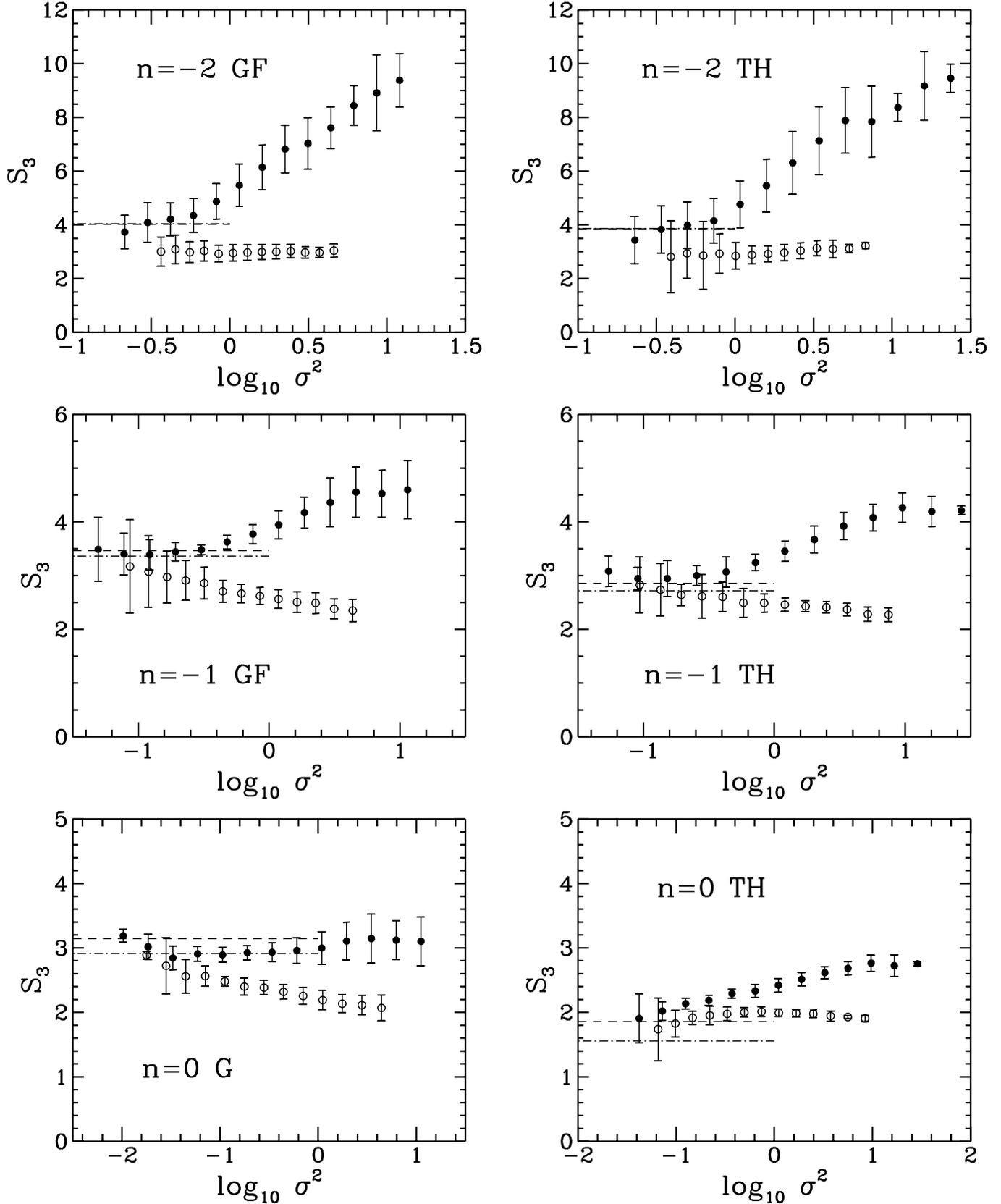

**Fig. 5.** Skewness factor in real space (filled circles) and redshift space (open circles) measured in $N$–body simulations. The dashes show the theoretically expected values in real space and the dots-dashes to the redshift space one, as given in Tables 2 and 3. The left panels correspond to a Gaussian smoothing and the right panels to a top-hat smoothing. The initial power-spectra considered here are scale invariant, with $n = -2$ (top), -1 (middle), and 0 (bottom).

for a top–hat filter. The quantity $\Delta \log_{10} \ell$ corresponds to the logarithmic scale step we chose. We used several snapshots of each simulation, corresponding to the various expansion factors $a$ (with initially $a = 1$) given in Table 4. For each snapshot, the constraint (51) is approximately fulfilled and $\ell_0$ remains reasonably small (as compared to $L_{\text{box}}$, Table 4). With such values of $\ell_0$ we expect finite volume effects to be small and we do not correct for them. However, in order to get an idea of the uncertainties related to the possible artifacts listed above, we tried to compute as realistic error–bars as possible. In particular, we used the procedure proposed by CBS to estimate finite volume effect errors (see Appendix C, where we discuss in detail the error–bars calculations).

**Table 4.** The expansion factor $a$ and the measured correlation length $\ell_0$ corresponding to the various snapshots chosen to measure the skewness in our $N$–body simulations

| $n$ | $a$ | $\ell_0$ |
|---|---|---|
|    | 2.5 | 0.027 |
| -2 | 3.2 | 0.040 |
|    | 4.0 | 0.054 |
|    | 2.5 | 0.015 |
| -1 | 4.0 | 0.022 |
|    | 6.4 | 0.036 |
|    | 4.1 | 0.011 |
| 0  | 8.1 | 0.018 |
|    | 16  | 0.028 |

4.2.2. Measurement results

Figure 5 displays the measured values of $S_3$ in real (filled circles) and redshift space (open symbols) as functions of the logarithm of the variance. Left panels correspond to Gaussian smoothing and right panels to top–hat smoothing. The index $n$ of the initial power spectrum increases from top to bottom. The dashed and dotted-dashed lines represent the analytical predictions of second order perturbation theory in real and redshift space respectively. They should superpose to the filled circles and the open circles in the weakly non–linear regime, i.e., when $\sigma^2 \ll 1$.

In real space, there is a good agreement between theory and measurement in the regime $\sigma \lesssim 1$, which confirms the results of previous authors. In redshift space, the agreement between the second order predictions and the measurement is globally less good (except for the case $n = -1$ with top–hat smoothing), unless $\sigma_z$ is much smaller than unity, typically $\sigma_z^2 \lesssim 0.1$. Indeed, the measured quantity $S_{3,z}$ is seen to be considerably smaller than $S_3$, particularly in the strongly non–linear regime. The $S_{3,z}$ vs. $\sigma^2$ curve tends also to be flatter than its real space image, $S_3(\sigma^2)$, as already noticed by Lahav et al. (1993); the overall deviation from a scale–invariant behavior $[S_3(\sigma^2) = \text{constant}]$ is also much less significant in redshift space than in real space. This deviation, which seems in real space to increase with decreasing $n$ (in agreement with the results of Lucchin et al. 1994) is particularly large for $n = -2$. In the latter case, the measured value of $S_3$ is amplified by more than a factor two as compared to the value seen in the weakly non–linear regime (left lower corners of top panels in Figure 5). Note however that the case $n = -2$ is expected to be strongly contaminated by finite volume effects (as already discussed by CBS and Lucchin et al. 1994). This is why the error–bars are so large in that case. Furthermore, the measurements of $S_3$ and $S_{3,z}$ are likely to be underestimates of the true value (See Appendix C and CBS).

The fact that $S_{3,z}$ is much smaller and flatter than $S_3$ in the non linear regime is certainly related to the "finger of god" effect described in the introduction. This is of course a highly non–linear effect. It has been analytically studied for the three-body correlation function by Matsubara (1994). His results, when generalized to averaged correlations should match our measurements. Furthermore, it is likely that this finger of god effect is also responsible for the slow convergence toward the perturbative predictions when the variance is decreased. Indeed, it smears dense structures along the line–of–sight up to fairly large scales. This behavior has already been observed in studies of redshift distortions of the power spectrum (e.g. Cole et al. 1994).

## 5. Conclusions

In this paper, we have studied redshift distortions of clustering in the weakly non–linear regime of the gravitational instability. Second–order Lagrangian perturbation theory was applied to spatially flat cosmological models with $\Omega + \Lambda/3H^2 = 1$ and to open models with $\Lambda = 0$. We have analytically estimated the bispectrum and the redshift space value $S_{3,z}$ of the skewness factor $S_3 \equiv \langle \delta^3 \rangle / \langle \delta^2 \rangle^2$ for both the unsmoothed and smoothed density field. We considered the case of Gaussian and top–hat smoothing for scale–invariant initial power-spectra $\langle |\delta_k|^2 \rangle \propto k^n$. We also compared our analytical results to measurements in $N$–body experiments (for $\Omega = 1$). The main results are the following:

1. The skewness factor and the $Q$–amplitude of the bispectrum are seen to be relatively insensitive to projection in redshift space in the weakly non–linear regime. In other words, the third moment and the square of the second moment are globally changed by nearly the same factor [$\sim b_1^2$, Eq. (33)], both in real space and redshift space. The most important differences are found for $\Omega = 1$, where $|S_{3,z} - S_3|/S_3 \lesssim 0.2$, which is small in view of the present uncertainties in the observationnal determinations of $S_3$.
2. The skewness, which was already known to be almost independent of the value of the density parameter $\Omega$ (for $\Lambda = 0$), depends only very weakly on the value of $\Lambda$ in a flat universe ($\Omega + \Lambda/3H^2 = 1$). The same result is valid for the bispectrum.
3. Analytical predictions for $S_3$ are in good agreement with measurements in $N$–body simulations but the finger of god effect is felt till quite large scales. Indeed, our measurements suggest that the $N$–body values of $S_{3,z}$ do converge to the analytical predictions only in the regime $\sigma^2 \lesssim 0.1$ (with $\sigma^2 \equiv \langle \delta^2 \rangle$). In real space, the agreement between theory and measurements was quite good for $\sigma$ almost as large as unity.

Peebles (1980) had shown that $S_3$ for the unsmoothed field has a specific value, 34/7, if large scale structures arose from the gravitational development of Gaussian initial conditions in an $\Omega = 1$, $\Lambda = 0$, model. Bouchet et al. (1992) showed that the effect of $\Omega$ and of the real space-redshift space mapping are quite weak. Juszkiewicz and Bouchet (1991) and Juszkiewicz et al. (1993a) computed the power spectrum dependence introduced by the smoothing of the evolved density field. Juszkiewicz et al. (1993b), and Fry and Gaztañaga (1993) analyzed the effect of a non-linear biasing. This paper completes the series by an analysis of the redshift space value of $S_3$ for a smoothed field in a large class of cosmological models. At all steps the results were carefully checked with simulations. Comparisons with redshift space measurements like those in the IRAS 1.2 Jy catalog (Bouchet et al. 1991, 1993) can then be used to test the hypothesis that the presently observed large scale structure was generated by gravitational instability, acting on initially Gaussian density fluctuations.

The real space expression for the bispectrum was first computed by Fry (1984), while the region of validity of this perturbative result was tested in numerical simulations by Fry et al. (1993). Most recently, Fry (1994) pointed out that measurements of the bispectrum can be used to measure the value of the non-linear bias. Here also we complete the series by analyzing how $\Omega$ and $\Lambda$ affect the bispectrum and the $Q$ parameter, measured in redshift space. As a result, it may for the first time become possible to use redshift surveys to estimate the linear and non–linear bias factors, breaking the coupling between the bias and cosmological parameters – $\Omega$ and $\Lambda$.

*Acknowledgements.* Part of this work was done while FRB was visiting Fermilab and while SC was at the Institut d'Astrophysique de Paris (CNRS), supported by Ecole Polytechnique. SC is now supported by DOE and by NASA through grant NAGW-2381 at Fermilab. RJ is supported by the Polish Committee for Scientific Research (KBN), grant No. 2-1243-91-01. RJ also thanks Alain Omont and John Bahcall for their hospitality respectively at the Institut d'Astrophysique de Paris and at the Institute for Advanced Study. The computational means (CRAY-98) were made available to us thanks to the scientific council of the Institut du Développement et des Resources en Informatique Scientifique (IDRIS). Finally, we have benefited from discussions at the "Gravitational Clustering in Cosmology" workshop at the Aspen Center for Physics.

## APPENDIX

### A. From real space to redshift space

Let us consider a spherical coordinate system centered on the observer. The mapping from real space to redshift space can then be written as

$$\boldsymbol{r} = a\boldsymbol{x} = (r, \theta, \varphi) \longrightarrow \boldsymbol{z} = (z, \theta, \varphi), \tag{A1}$$

where $z$ is the "comoving" redshift given by $z \equiv V_r/(da/dt)$ and $V_r$ is the proper velocity projected along the line–of–sight. In the Lagrangian approach, the comoving coordinate of a fluid element (a "particle") is

$$\boldsymbol{x}(t) = \boldsymbol{q} + \varepsilon \boldsymbol{\Psi}^{(1)}(\boldsymbol{q}, t) + \varepsilon^2 \boldsymbol{\Psi}^{(2)}(\boldsymbol{q}, t) + \mathcal{O}(\varepsilon^3). \tag{A2}$$

We assume here that the fastest growing modes, $g_i(t)$, $i = 1, 2$, are dominating. We can thus write $\boldsymbol{\Psi}^{(i)}(\boldsymbol{q}, t) = g_i(t)\tilde{\boldsymbol{\Psi}}^{(i)}(\boldsymbol{q}, t)$. The comoving redshift coordinate of the element of matter is then

$$\begin{aligned}z(\boldsymbol{q}) &= q_x + \varepsilon(1 + f_1)\Psi_x^{(1)}(\boldsymbol{q}, t) \\ &+ \varepsilon^2(1 + f_2)\Psi_x^{(2)}(\boldsymbol{q}, t) + \mathcal{O}(\varepsilon^3),\end{aligned} \tag{A3}$$

with $f_i = (a/g_i)(dg_i/da)$. We also denote by a subscript the radial component of a vector, i.e. $u_x$ stands for the radial component of the vector $\boldsymbol{u}$, $u_x \equiv \boldsymbol{u} \cdot \boldsymbol{x}/x = \boldsymbol{u} \cdot \boldsymbol{r}/r$.

By introducing the new vectors $\bar{\boldsymbol{\Psi}}^{(1)}(\boldsymbol{x}, t) = \boldsymbol{\Psi}^{(1)}(\boldsymbol{q} = \boldsymbol{x}, t)$ and $\bar{\boldsymbol{\Psi}}^{(2)}(\boldsymbol{x}, t) = \boldsymbol{\Psi}^{(2)}(\boldsymbol{q} = \boldsymbol{x}, t)$ which obey equations (9), (10), but in **x** space instead of **q** space, we can express $z$ in Eulerian coordinates as

$$\begin{aligned}z(\boldsymbol{x}) &= x + \varepsilon f_1 \bar{\Psi}_x^{(1)}(\boldsymbol{x}) \\ &+ \varepsilon^2 \left( f_2 \bar{\Psi}_x^{(2)}(\boldsymbol{x}) - f_1 \bar{\boldsymbol{\Psi}}^{(1)} . \nabla_x \bar{\Psi}_x^{(1)} \right) + \mathcal{O}(\varepsilon^3).\end{aligned} \tag{A4}$$

### A.1. density contrast

We now write the density contrast in redshift space, as a function of $\boldsymbol{z}$. In redshift space, the density contrast $\delta_z \equiv (\rho_z - \bar{\rho})/\bar{\rho}$ around a particle initially in $\boldsymbol{q}$ is given by the requirement of mass conservation:

$$\delta_z(\boldsymbol{q}) = \left|\frac{\partial \boldsymbol{z}}{\partial \boldsymbol{q}}\right|^{-1} - 1. \tag{A5}$$

The direct use of Eq. (A5) is however a little awkward, because the vectors $\boldsymbol{z}$ and $\boldsymbol{q}$ are not necessarily collinear. We found simpler and more illuminating to first compute the density contrast in real space as a function of the Eulerian coordinate $\boldsymbol{x}$, using mass conservation requirement (5) and then pass to redshift space using the transformation (A4). The development at second order in $\varepsilon$ of Eq. (5) is

$$\begin{aligned}\delta_x(\boldsymbol{q}) &= -\varepsilon \nabla_q . \boldsymbol{\Psi}^{(1)} \\ &+ \varepsilon^2 \left\{ (\nabla_q . \boldsymbol{\Psi}^{(1)})^2 + \left(1 + \frac{g_1^2}{g_2}\right) \nabla_q . \boldsymbol{\Psi}^{(2)} \right\}.\end{aligned} \tag{A6}$$

Using the Eulerian displacements $\bar{\boldsymbol{\Psi}}^{(i)}(\boldsymbol{x}, t)$ defined above and the following property

$$\nabla_q . \boldsymbol{\Psi}^{(1)} = \nabla_x . \bar{\boldsymbol{\Psi}}^{(1)} - \varepsilon \left\{ \nabla_x (\nabla_x . \bar{\boldsymbol{\Psi}}^{(1)}) \right\} \bar{\boldsymbol{\Psi}}^{(1)} + \mathcal{O}(\varepsilon^2), \tag{A7}$$

we can easily get $\delta_x$ as a function of $\boldsymbol{x}$ at second order in $\varepsilon$ (e.g. Bouchet et al. 1994). Then, using the fact that $\boldsymbol{z} \propto \boldsymbol{x}$, we simply write the mapping from real space to redshift space as

$$\frac{\rho_z}{\rho_x} = \frac{x^2}{z^2}\left|\frac{\partial x}{\partial z}\right|. \tag{A8}$$

By using Eq. (A4), we are now able to write the density contrast in redshift space as a function of $x$:

$$\rho_z(x) = 1 - \varepsilon \left( \nabla \cdot \bar{\Psi}^{(1)} + f_1 \frac{\partial}{x^2 \partial x}(x^2 \bar{\Psi}_x^{(1)}) \right)$$
$$+ \varepsilon^2 \left[ (\nabla \cdot \bar{\Psi}^{(1)})^2 + \bar{\Psi}^{(1)} \cdot \nabla(\nabla \cdot \bar{\Psi}^{(1)}) - (1 + \frac{g_1^2}{g_2})\nabla \cdot \bar{\Psi}^{(2)} \right.$$
$$+ f_1 \left( \frac{\partial}{x^2 \partial x}(x^2 \bar{\Psi}_x^{(1)}) \nabla \cdot \bar{\Psi}^{(1)} + \frac{\partial}{x^2 \partial x}(x^2 \bar{\Psi}^{(1)} \cdot \nabla \bar{\Psi}_x^{(1)}) \right)$$
$$+ f_1^2 \left( \frac{\partial}{x^2 \partial x}(x^2 \bar{\Psi}_x^{(1)}) \frac{\partial}{\partial x} \bar{\Psi}_x^{(1)} + 3 \frac{\Psi_x^{(1)\,2}}{x^2} \right)$$
$$\left. - f_2 \frac{\partial}{x^2 \partial r}(x^2 \bar{\Psi}_x^{(2)}) \right]. \quad (A9)$$

The last step is to express $\delta_z$ as a function of $z$. We introduce the new vectors $\hat{\Psi}^{(1)}(z,t) = \bar{\Psi}^{(1)}(x=z,t)$ and $\hat{\Psi}^{(2)}(z,t) = \bar{\Psi}^{(2)}(x=z,t)$. These displacements obey the equations (9) and (10), but with the calculations done in $z$ space. Using the following properties,

$$\nabla_x \cdot \bar{\Psi}^{(1)} = \nabla_z \cdot \hat{\Psi}^{(1)} - \varepsilon f_1 \hat{\Psi}_z^{(1)} \frac{\partial}{\partial z}(\nabla_z \cdot \hat{\Psi}^{(1)}), \quad (A10)$$

$$\frac{\partial}{x^2 \partial x}(x^2 \bar{\Psi}_x^{(1)}) = \frac{\partial}{z^2 \partial z}(z^2 \hat{\Psi}_z^{(1)})$$
$$- \varepsilon \hat{\Psi}_z^{(1)} \frac{\partial}{\partial z} \left( \frac{\partial}{z^2 \partial z}(z^2 \hat{\Psi}_z^{(1)}) \right), \quad (A11)$$

we obtain the density contrast in redshift space as a function of $z$, as given in Eqs. (20), (21), (23), (24) and (25). In the following, we use the "large volume approximation" introduced in Sect. 2.2, which results in a considerable simplification of the expression of $\delta_z(z)$ [i.e. Eqs. (27), (28)].

The second and third moments of the density contrast are computed by ensemble averaging:

$$\langle \delta^2(z) \rangle_z = \varepsilon^2 \langle \delta_1^2 + 2\delta_1 \Delta_1 + \Delta_1^2 \rangle \quad (A12)$$
$$\langle \delta^3(z) \rangle_z = \varepsilon^3 \langle (\delta_1 + \Delta_1)^3 \rangle + 3\varepsilon^4 \langle \delta_1^2 \delta_2 + \delta_1^2 \Delta_2$$
$$+ 2\delta_1 \Delta_1 \delta_2 + 2\delta_1 \Delta_1 \Delta_2 + \Delta_1^2 \delta_2 + \Delta_1^2 \Delta_2 \rangle \quad (A13)$$

In order to consider the more general case of a density contrast field smoothed with a window $W(z)$, $\delta_z(z)$ in Eq. (20) has to be replaced by $\delta_z^S(z) = \delta_z * W(z)$.

### A.2. Fourier Transforms

For a spherically symmetric filter, the smoothed density contrast is given by

$$\delta_z^S(z) = \sum_k \delta_z(\mathbf{k}) W(k), \quad (A14)$$

and the displacement field is

$$\Psi^{(1)}(z,t) = g_1(t) \sum_k \Psi_k^{(1)} e^{i\mathbf{k} \cdot z}. \quad (A15)$$

Since

$$\delta_1 = -\nabla_z \cdot \Psi^{(1)}(z) = -g_1(t) \sum_k \hat{\delta}_1(\mathbf{k}) e^{i\mathbf{k} \cdot z}, \quad (A16)$$

we get

$$\Psi_k^{(1)} = \frac{\mathbf{k}}{ik^2} \hat{\delta}_1(\mathbf{k}), \quad (A17)$$

and

$$\Delta_1(z,t) = g_1(t) \sum_k \hat{\delta}_1(\mathbf{k}) W(k) f_1 \mu_k^2 e^{i\mathbf{k} \cdot z}, \quad (A18)$$

$$\delta_2(z,t) = g_1^2(t) \sum_{k,l} \hat{\delta}_1(\mathbf{k}) \hat{\delta}_1(\mathbf{l}) W(|\mathbf{k}+\mathbf{l}|)$$
$$\left(1 + \cos\theta \frac{k}{l} - (1 + \frac{g_2}{g_1^2})\frac{1-\cos^2\theta}{2} \right) e^{i(\mathbf{k}+\mathbf{l}) \cdot z} \quad (A19)$$

$$\Delta_2(z) = g_1^2(t) \sum_{k,l} \hat{\delta}_1(\mathbf{k}) \hat{\delta}_1(\mathbf{l}) W(|\mathbf{k}+\mathbf{l}|)(k\mu_k + l\mu_l)$$
$$\left( f_1 \frac{\mu_k}{k}(1 + \cos\theta \frac{k}{l}) + f_1^2 \frac{\mu_k}{k} \mu_l^2 \right.$$
$$\left. - f_2 \frac{g_2}{g_1^2} \frac{1-\cos^2\theta}{2} \frac{(k\mu_k + l\mu_l)}{(\mathbf{k}+\mathbf{l})^2} \right) e^{i(\mathbf{k}+\mathbf{l}) \cdot z}, \quad (A20)$$

with

$$\mu_k \equiv \frac{\mathbf{k} \cdot z}{kz}, \quad \mu_l \equiv \frac{\mathbf{l} \cdot z}{lz}, \quad \cos\theta \equiv \frac{\mathbf{k} \cdot \mathbf{l}}{kl}.$$

In this paper, we assume that the linear density contrast, $\delta_1$, is Gaussian. Its Fourier coefficients, $g_1 \hat{\delta}_i$, are thus uncorrelated and satisfy Eqs. (29), (30), (31). This implies for instance [Eq. (30)] that the first term of (A13) vanishes. The first non trivial term of $\langle \delta^3 \rangle$ is then

$$\langle \delta_1^2 \delta_2 \rangle = \iint \frac{d\mathbf{k}\, d\mathbf{l}}{(2\pi)^6} P(k) P(l) W(k) W(l) W(|\mathbf{k}+\mathbf{l}|)$$
$$\left\{ 2 + \cos\theta \left( \frac{k}{l} + \frac{l}{k} \right) - \left( 1 + \frac{g_2}{g_1^2} \right)(1 - \cos^2\theta) \right\} \quad (A21)$$

by using the equations (A13), (A18-A20), (31). In the same way, all terms of the second and third moments simplify thanks to Eq. (31). The averaging over $z$ of $\mu_k$ and $\mu_l$ factors yields functions of $\cos\theta$. Eventually one gets:

$$\langle \delta^2 \rangle = b_1 \varepsilon^2 g_1^2 \int \frac{d\mathbf{k}}{(2\pi)^3} P(k) W^2(k) \quad (A22)$$

$$\langle \delta^3 \rangle = 3\varepsilon^4 g_1^4 \iint \frac{d\mathbf{k}\, d\mathbf{l}}{(2\pi)^6} P(k) P(l) W(k) W(l) W(|\mathbf{k}+\mathbf{l}|)$$
$$\left( a_0 + a_2 \cos^2\theta + a_4 \cos^4\theta \right.$$
$$+ \left( a_1 \cos\theta + a_3 \cos^3\theta \right) \left( \frac{k}{l} + \frac{l}{k} \right)$$
$$\left. + (1-\cos^2\theta)^2 \left( \frac{x_1(k^2+l^2) + x_2\, kl \cos\theta}{k^2+l^2+2kl\cos\theta} \right) \right), (A23)$$

where the factors $b_1$, $a_i$ and $x_j$ are listed in Sect. 3 [Eqs. (33), (35)].

Note that by setting $f_1 = f_2 = 0$ in the expressions above, we can recover the real space values of all the quantities just computed. JB and JBC had developed techniques to compute the skewness of the smoothed density field in real space. These techniques will now be described in detail and applied to the case of redshift space. The real space results will also be presented.

## B. Effect of smoothing

We now compute the skewness of the density field with either no smoothing, a smoothing with a Gaussian filter, or smoothing with a top-hat filter. We deal only with power law models, $P(k) = A k^n$. As these models are scale-free, we can set the smoothing length to unity without loss of generality.

### B.1. Unsmoothed Result

In the case without any smoothing ($W(k) \equiv 1$), the integration of the equations (A22), (A23) over $k$ and $l$ in spherical coordinates yields

$$\frac{\langle \delta^3 \rangle}{\langle \delta^2 \rangle^2} = \frac{3}{b_1^2} \left( a_0 + \frac{a_2}{3} + \frac{a_4}{5} + \frac{1/2}{\left( \int_0^\infty dk\, k^2\, P(k) \right)^2} \int_0^\infty dk\, k^2\, P(k) \int_0^\infty dl\, l^2\, P(l) \int_{-1}^1 du\, X(k, l, u) \right) \quad \text{(B1)}$$

with $u = \cos\theta$. The last term

$$X(k, l, u) = \frac{(1 - u^2)^2 (x_1 (k^2 + l^2) + x_2\, klu)}{(k^2 + l^2 + 2\, klu)} \quad \text{(B2)}$$

is problematic because $k$, $l$, and $u$ cannot be separated. However, thanks to the inequalities $k^2 + l^2 > 2kl > 0$, it is easy to bound this term between polynomials of $u$. Integrating these polynomials, we obtain

$$\frac{29}{20} x_1 - \frac{23}{120} x_2 < \int_{-1}^1 du\, X(k, l, u) < \frac{19}{20} x_1 + \frac{7}{120} x_2.$$

Some numerical values are given in Sect. 4.1.

### B.2. Smoothing with a Gaussian filter

We now consider a density contrast field filtered with a Gaussian window of variance unity and normalized to unity. Its Fourier transform is then $W(k) = \exp(-k^2/2)$. The integration over the angular variables of $k$ and $l$ in the equations (A22) and (A23) yields

$$\langle \delta^2 \rangle = \varepsilon^2 \frac{b_1 A}{2\pi^2} \int_0^\infty dx\, x^{n+2} \exp^{-x^2}, \quad \text{(B3)}$$

$$\langle \delta^3 \rangle = \varepsilon^4 \frac{3 A^2}{8 \pi^4} \int_0^\infty dx \int_0^\infty dy \int_{-1}^1 du\, x^{n+2} y^{n+2} \exp^{-(x^2 + y^2 + xyu)}$$

$$\left( a_0 + a_2\, u^2 + a_4\, u^4 + \left( a_1\, u + a_3\, u^3 \right) \frac{2y}{x} + (1 - u^2)^2 \left( \frac{x_1\, (x^2 + y^2) + x_2\, xyu}{x^2 + y^2 + 2\, xyu} \right) \right). \quad \text{(B4)}$$

The case $n = -3$ is special and must be treated separately from the case of the other $n$.

### B.2.1. Case when $n = -3$

When the power spectrum index $n$ is equal to $-3$, both $\langle \delta^2 \rangle$ and $\langle \delta^3 \rangle$ diverge. But we can still give a meaning to $S_3$, if we define it as its value in the limit, if any, when $n$ tends to $-3$. By introducing $\epsilon = n + 3$, one gets

$$\langle \delta^2 \rangle = \frac{b_1 A}{4\pi^2} \Gamma(\frac{\epsilon}{2}),$$

where $\Gamma$ is the Euler Gamma function, which behaves as $1/\epsilon$ when $\epsilon$ goes to zero. For $\langle \delta^3 \rangle$, it is convenient to change variables, from the Cartesian $(x, y)$ to the polar ones $(\rho, \varphi)$ defined by $x = \rho \cos\varphi$, $y = \rho \sin\varphi$. The integration over $\rho$ from 0 to $\infty$ then yields

$$\langle \delta^3 \rangle = \frac{3 A^2}{16\pi^4} \int_{-1}^1 du \int_0^{\pi/2} d\varphi \frac{\Gamma(\epsilon)}{(1 + \frac{u}{2} \sin 2\varphi)^\epsilon} \cos^{\epsilon - 1}\varphi \sin^{\epsilon - 1}\varphi \left( a_0 + a_2 u^2 + a_4 u^4 + (1 - u^2)^2 \frac{(x_1 + u x_2)}{1 + u \sin 2\varphi} + 2(a_1 u + a_3 u^3) \tan\varphi \right). \quad \text{(B5)}$$

The integrand of Eq. (B5) includes the terms:

$$(1 + \frac{u}{2} \sin 2\varphi)^{-\epsilon} = 1 - \epsilon \log(1 + \frac{u}{2} \sin 2\varphi) + \mathcal{O}(\epsilon^2)$$

which always converges since $|u/2 \sin 2\varphi| < 1$, and

$$\frac{\cos^{\epsilon - 1}\varphi \sin^{\epsilon - 1}\varphi}{1 + u \sin 2\varphi} = -u \frac{2^{1-\epsilon} \sin^\epsilon 2\varphi}{1 + u \sin 2\varphi} + \cos^{\epsilon - 1}\varphi \sin^{\epsilon - 1}\varphi.$$

In this last equation, the first term on the right hand side is bounded as $\epsilon$ tends to 0, whereas the second one verifies

$$\int_0^{\pi/2} d\varphi \cos^{\epsilon - 1}\varphi \sin^{\epsilon - 1}\varphi = \frac{\Gamma^2(\epsilon/2)}{2\Gamma(\epsilon)}$$

(e.g. Gradshteyn & Ryzhik 3.621.5). Integrals of terms with an odd degree in $u$ vanish and one gets for the skewness

$$S_3(n = -3) = \frac{3}{b_1^2} \left( a_0 + \frac{a_2}{3} + \frac{a_4}{5} + \frac{16}{15} \frac{x_1}{2} \right).$$

This yields for instance

$$\Omega = 1, \quad S_3 = 5.0217,$$

$$\Omega = .1, \quad S_3 = 4.9469.$$

### B.2.2. Case when $n$ is not $-3$

The term

$$X(x, y, u) = (1 - u^2)^2 \left( \frac{x_1 \, (x^2 + y^2) + x_2 \, xyu}{x^2 + y^2 + 2 \, xyu} \right)$$

in the Eq. (B4) makes it impossible to integrate analytically $\langle \delta^3 \rangle$. Fortunately, this term can be bounded. To begin with, let us suppose that it is a constant, and let us set $u = 2\cos\beta$. Equation (B4) then becomes

$$\begin{aligned}\langle \delta^3 \rangle &= \varepsilon^4 \frac{3A^2}{4\pi^4} \int_{\pi/3}^{2\pi/3} d\beta \sin\beta \\ &\quad (G_n(\beta) \left(X + a_0 + 4 a_2 \cos^2\beta + 16 a_4 \cos^4\beta\right) \\ &\quad + F_n(\beta) \left(4 a_1 \cos\beta + 16 a_3 \cos^3\beta\right)) , \end{aligned} \quad \text{(B6)}$$

with

$$F_n(\beta) \equiv \int_0^\infty dx \int_0^\infty dy \, x^{n+1} y^{n+3} e^{-x^2 - y^2 - 2xy\cos\beta}, \quad \text{(B7)}$$

$$G_n(\beta) \equiv \int_0^\infty dx \int_0^\infty dy \, x^{n+2} y^{n+2} e^{-x^2 - y^2 - 2xy\cos\beta}. \quad \text{(B8)}$$

We can now evaluate $F_{-1}(\beta)$, by introducing the new variables $v$ and $w$ defined by

$$x = v - w\cot\beta, \quad y = \frac{w}{\sin\beta}, \quad \left|\frac{\partial(x,y)}{\partial(v,w)}\right| = \frac{1}{\sin\beta}.$$

The integral becomes

$$F_{-1}(\beta) = \frac{1}{\sin^3\beta} \int_0^\infty dw \int_{w\cot\beta}^\infty dv \, e^{-v^2 - w^2} w^2.$$

In the polar coordinates $v = \rho\cos\varphi$, $w = \rho\sin\varphi$, the condition $w \geq 0$ gives $0 \leq \varphi \leq \pi$, while $v \geq w\cot\beta$ leads to $\cot\varphi \geq \cot\beta$, or $0 \leq \varphi \leq \beta$. The integral is then given by

$$F_{-1}(\beta) = \frac{1}{4} \left( \frac{\beta}{\sin^3\beta} - \frac{\cos\beta}{\sin^2\beta} \right).$$

Similar techniques can be used to evaluate

$$G_{-1}(\beta) = \frac{1}{4} \left( \frac{1}{\sin^2\beta} - \frac{\beta\cos\beta}{\sin^3\beta} \right).$$

From the definitions (B7) and (B8), one can express $F_n$ and $G_n$, for any value of $n > -1$, as

$$F_n(\beta) = (-2)^{-n-1} \frac{d^{n+1}}{d\cos^{n+1}\beta} F_{-1}(\beta),$$

$$G_n(\beta) = (-2)^{-n-1} \frac{d^{n+1}}{d\cos^{n+1}\beta} G_{-1}(\beta).$$

Similarly, $F_{-2}$ and $G_{-2}$ are integrals over $\beta$ of $2\sin\beta$ times $F_{-1}$ and $G_{-1}$ respectively

$$F_{-2}(\beta) = -\frac{1}{2}\beta\cot\beta, \quad G_{-2}(\beta) = \frac{\beta}{2\sin\beta}.$$

Equation (B6) can be integrated, and one obtains an interval for $S_3$ according to the computed bounds on the term $X$. The three-dimensional integral of $X$ is first reduced to a one-dimensional integral of an impressively long function which happens to be very smooth. This function can then be integrated numerically to yield an exact value of the skewness. Numerical values of the skewness for different power spectra in real and redshift space are listed in Table 2.

### B.3. Smoothing with a top-hat filter

We now consider a top-hat filter of radius unity and normalized to unity. Its Fourier transform is then

$$W(k) = 3k^{-3}(\sin k - k\cos k) = 3\sqrt{\frac{\pi}{2}} k^{-3/2} J_{3/2}(k),$$

where $J_p(x)$ denotes the spherical Bessel function of order $p$. We decompose equation (A23) in two terms according to

$$\langle \delta^3 \rangle = \varepsilon^4 \left(I_{\text{t.h.}} + H_{\text{t.h.}}\right)$$

where $I_{\text{t.h.}}$ can be computed analytically and $H_{\text{t.h.}}$ can not. That is

$$\begin{aligned}I_{\text{t.h.}} &\equiv 3 \iint \frac{d\mathbf{k}\, d\mathbf{l}}{(2\pi)^6} k^n l^n \left(3\sqrt{\frac{\pi}{2}}\right)^3 k^{-3/2} J_{3/2}(k) \\ &\quad l^{-3/2} J_{3/2}(l) |\mathbf{k}+\mathbf{l}|^{-3/2} J_{3/2}(|\mathbf{k}+\mathbf{l}|) \left(a_0 + a_2 \cos^2\theta \right. \\ &\quad \left. + a_4 \cos^4\theta + (a_1\cos\theta + a_3\cos^3\theta)\left(\frac{k}{l} + \frac{l}{k}\right)\right),\end{aligned}$$

$$\begin{aligned}H_{\text{t.h.}} &\equiv 3 \iint \frac{d\mathbf{k}\, d\mathbf{l}}{(2\pi)^6} k^n l^n \left(3\sqrt{\frac{\pi}{2}}\right)^3 \\ &\quad k^{-3/2} J_{3/2}(k)\, l^{-3/2} J_{3/2}(l)\, |\mathbf{k}+\mathbf{l}|^{-3/2} J_{3/2}(|\mathbf{k}+\mathbf{l}|) \\ &\quad \left((1-\cos^2\theta)^2 \frac{x_1 \, (k^2 + l^2) + x_2 \, kl\cos\theta}{k^2 + l^2 + 2\, kl\cos\theta}\right).\end{aligned}$$

By introducing the Legendre Polynomials $P_m$, and with the change of variable $\mathbf{l} \to -\mathbf{l}$ justified further down, $I_{\text{t.h.}}$ becomes

$$\begin{aligned}I_{\text{t.h.}} &= \frac{27}{16\pi^3}\sqrt{\frac{\pi}{2}} \int_0^\infty Dk \int_0^\infty dl \int_{-1}^1 du \, k^{n+2} l^{n+2} \\ &\quad k^{-3/2} J_{3/2}(k)\, l^{-3/2} J_{3/2}(l)\, w^{-3/2} J_{3/2}(w) \\ &\quad (\alpha_0 P_0(u) + \alpha_2 P_2(u) + \alpha_4 P_4(u) \\ &\quad - \frac{k}{l} (\alpha_1 P_1(u) + \alpha_3 P_3(u))). \end{aligned} \quad \text{(B9)}$$

We have used

$$u = \cos(\mathbf{k}, \mathbf{l}), \quad w = |\mathbf{k} - \mathbf{l}|,$$

and

$$\alpha_0 = 3(a_0 + \frac{a_2}{3} + \frac{a_4}{5}), \quad \alpha_1 = 6(a_1 + \frac{3}{5}a_3),$$

$$\alpha_2 = 3(\frac{2}{3}a_2 + \frac{4}{7}a_4), \quad \alpha_3 = \frac{12}{5}a_3, \quad \alpha_4 = \frac{24}{35}a_4.$$

The term $J_{3/2}(w)$ is clearly a source of trouble because it mixes all three variables $k$, $l$ and $u$. Fortunately one can use Gegenbauer's addition theorem, which states for $w = \sqrt{k^2 + l^2 - 2klu}$ (e.g. Watson 1958):

$$w^{-3/2} J_{3/2}(w) = \sqrt{2\pi} \sum_{m=0}^{\infty} (\frac{3}{2} + m)(kl)^{-3/2}$$
$$J_{3/2+m}(k) \, J_{3/2+m}(l) \, \frac{d}{du} P_{m+1}(u). \quad (B10)$$

Equation (B9) can now be expanded into an infinite series of products of integrals over $k$, $l$ and $u$. Integral over $u$ can be simplified using orthogonality properties of Legendre polynomials, to yield:

$$\sum_{m=0}^{\infty} g(m) \int_{-1}^{1} du \, P_q(u) \frac{d}{du} P_{m+1}(u) = 2 \sum_{m=0}^{\infty} g(2m + q).$$

Let us define $\mathcal{F}(\nu, \mu, -\lambda) \equiv \int_0^{\infty} J_\nu(x) J_\mu(x) x^{-\lambda} dx$. It verifies (Gradshteyn & Ryzhik 6.574.2)

$$\mathcal{F}(\nu, \mu, -\lambda) = \frac{\Gamma(\lambda) \, \Gamma(\frac{\nu+\mu-\lambda+1}{2})}{2^\lambda \, \Gamma(\frac{-\nu+\mu+\lambda+1}{2}) \Gamma(\frac{\nu+\mu+\lambda+1}{2}) \Gamma(\frac{\nu-\mu+\lambda+1}{2})} \quad (B11)$$

for $\nu + \mu + 1 > \lambda > 0$. The second moment then is

$$\langle \delta^2 \rangle = \frac{9}{4\pi} b_1 \mathcal{F}(\frac{3}{2}, \frac{3}{2}, n - 1),$$

and is finite for $-3 < n < 1$. And the skewness is

$$\frac{I_{\text{t.h.}}}{\langle \delta^2 \rangle^2} = \frac{1}{3 b_1^2} \left[ 3 \alpha_0 - 5 \alpha_1 \frac{(1-n)(3+n)}{(2-n)(5-n)} \right.$$
$$+ \mathcal{R}_n(1)^2 \left( 7 (\alpha_0 + \alpha_2) - 9 (\alpha_1 + \alpha_3) \mathcal{Q}_n(1) \right)$$
$$+ \sum_{q=2}^{\infty} \mathcal{R}_n(q)^2 \left( (4q+3) (\alpha_0 + \alpha_2 + \alpha_4) \right.$$
$$\left. \left. - (4q+5) (\alpha_1 + \alpha_3) \mathcal{Q}_n(q) \right) \right],$$

with

$$\mathcal{R}_n(q) \equiv \frac{\mathcal{F}(3/2, 2q+3/2, n-1)}{\mathcal{F}(3/2, 3/2, n-1)},$$
$$\mathcal{Q}_n(q) \equiv \frac{\mathcal{F}(3/2, 2q+5/2, n) \, \mathcal{F}(3/2, 2q+5/2, n-2)}{\mathcal{F}(3/2, 2q+3/2, n-1)^2}.$$

With the help of Eq. (B11) and the identity

$$\frac{\Gamma(x)}{\Gamma(x-k)} = (-1)^k \frac{\Gamma(k+1-x)}{\Gamma(1-x)},$$

these ratios simplify to:

$$\mathcal{R}_n(q) = (-1)^q \frac{\Gamma(q + \frac{3+n}{2}) \Gamma(q + \frac{n}{2}) \Gamma(\frac{2-n}{2}) \Gamma(\frac{5-n}{2})}{\Gamma(\frac{3+n}{2}) \Gamma(\frac{n}{2}) \Gamma(q + \frac{2-n}{2}) \Gamma(q + \frac{5-n}{2})},$$
$$\mathcal{Q}_n(q) = \frac{(1-n)(2q+n)(2q+3+n)}{n(2q+2-n)(2q+5-n)}.$$

One can now identify some very simple special cases:

$$n = -3, \quad S_3^{\text{anal}} = \frac{\alpha_0}{b_1^2},$$
$$n = -2, \quad S_3^{\text{anal}} = \frac{1}{b_1^2}(\alpha_0 + \frac{\alpha_0 - \alpha_1 + \alpha_2}{84} - \frac{\alpha_1}{6}),$$
$$n = 0, \quad S_3^{\text{anal}} = \frac{1}{b_1^2}(\alpha_0 - \frac{\alpha_1}{2}).$$

For $n = 1$, $\mathcal{R}_1(q) = (-1)^q$, $\mathcal{Q}_1(q) = 0$, and the series diverges. For other values of $n$ the series has to be computed numerically and appears to converge very fast at least for $n \leq 0$.

We now have to compute the extra term $H$ in $\langle \delta^3 \rangle$. For $n = -3$, $\langle \delta^2 \rangle$ diverges and in the expansion of $H$ according to Eq. (B10) only the first term, with index $m = 0$, is divergent and contributes to skewness. One then gets $S_3 = b_1^{-2}(\alpha_0 + 8x_1/15)$. For all other values of $n$, $H$ has to be computed numerically. This time, the integral cannot be reduced or simplified, and must be calculated as a three-dimensional one. Because of the presence of Bessel functions in the integrand, it oscillates very rapidly and we chose to use a Monte–Carlo integration. According to the values of $x_1$ and $x_2$ (see Table) these integrals give a small contribution to the skewness and the numerical calculations could be done with a relative uncertainty that we estimate to be better than 1%. Thus, our numerical values of skewness are quite accurate, and the errors come mainly from the approximations we used for the logarithmic derivatives of the growth rates $f_1$ and $f_2$ when $\Omega \neq 1$.

### C. Measurement of $S_3$: error–bars

Here, we explain how we computed the error–bars shown in Fig. 5. Such errors are mainly due to finite volume effects and to the finite sampling of the measured density field. Note however that there will necessarily be some differences between various estimates of $S_3(\sigma^2)$ (corresponding to various snapshots in a simulation) at given $n$. One can estimate an average standard deviation associated to such variations

$$\Delta_{\text{SS}}^2 S_3 \sim \frac{1}{N_{\text{SS}} - 1} \sum_{\text{snapshot } S} \left( S_{3,S} - \langle S_3 \rangle \right)^2, \quad (C1)$$

where $N_{\text{SS}}$ is the total number of snapshots considered (three in our case) and $\langle S_3 \rangle$ is the average of the measured $S_3(\sigma^2)$ over all the snapshots ($S_{3,S}$ is of course the measured value of $S_3$ in the snapshot $S$). Although this error is in fact correlated with the finite volume error and the finite sampling errors which we discuss hereafter, it can also be due to other defects, such as those related to initial conditions. We shall consider such an error as independent from others.

To estimate the errors due to finite volume effects, we first fit the large-$\rho$ tail of the measured distribution function $P(\rho)$ by an exponential

$$P(\rho) = P(\rho_{\text{max}}) \exp(-\beta[\rho - \rho_{\text{max}}]), \quad (C2)$$

with

$$\beta = -\frac{1}{P(\rho_{\max})}\frac{dP}{d\rho}(\rho_{\max}). \tag{C3}$$

The error $\Delta_{\rm FV}$ due to finite volume effects on $S_3$, is always in the same direction, i.e., $S_3$ is always underestimated by direct measurements (CBS), since one misses the positive contribution due to the missing tail of the exponential. Following CBS, we use their Eq. (42) to write

$$\Delta_{\rm FV} S_3 \sim [\zeta^3 + 3\zeta^2 + 6\zeta + 6]\frac{P(\rho_{\max})}{\sigma^2 \beta^4}, \tag{C4}$$

with $\zeta \equiv \beta\rho_{\max}$. The fit is done using a least square fit method in the $(\rho, \log P)$ coordinate system (where an exponential looks like a straight line).

The smoothing of the density field can be done either in real space or in Fourier space. The smoothing in Fourier space is much simpler to do than in real space, and much faster in terms of computer calculation. However, in the case one works in redshift space, it is impossible, practically, to have a sample obeying periodic boundary conditions, which makes the Fourier space calculation difficult, because of edge effects. Such artifacts are expected to be small if the scale considered is small compared to the sample size, but it can become rather important (inducing errors larger than $\sim 10\%$) as soon as $\ell \gtrsim L_{\rm box}/20$. We thus decided to make all the calculations in real space, both for a top–hat smoothing and a Gaussian smoothing. In the case of a Gaussian smoothing, there is the added complication of the infinite spatial extent of the filter. One thus need to truncate the window at a finite radius $r$. We chose to make such a truncation at $r = T\ell$ with $T = 3$, i.e.,

$$\begin{aligned} W_\ell(r) &\propto \exp\left(-\frac{r^2}{2\ell^2}\right), & r \leq 3\ell, \\ W_\ell(r) &= 0, & r > 3\ell. \end{aligned} \tag{C5}$$

A comparison of the results obtained using the procedure above with those obtained in Fourier space (without redshift mapping) shows very little differences, much less than one percent. Moreover, we also checked that taking $T = 4$ instead of $T = 3$ did not significantly change the results.

Note that the calculations are much more costly for a Gaussian filter than for a top–hat one, because, at a given site, we have to take into account the contribution of the density field up to $3\ell$ in the first case and only up to $\ell$ in the second case. Hence, we did slightly less accurate calculations for a Gaussian smoothing than for a top–hat smoothing, in terms of number of sites sampled to evaluate the smoothed density distribution. In both cases, we computed the smoothed density field on a lattice with $N_{\rm s} \equiv N_{\rm g}^3$ sites, but $N_{\rm g}$ was smaller for a Gaussian smoothing than for a top–hat smoothing. To sample appropriately the smoothed density field with this method, $N_{\rm g}$ should be large enough that the distance between two sites is smaller than the smoothing scale $\ell$ considered. In other words, we should have $\ell \gtrsim L_{\rm box}/N_{\rm g}$, a condition fulfilled by our calculations. In order to evaluate the error associated with finite sampling, we used a different method for a Gaussian smoothing and for a top–hat smoothing. In the first case, we had an integral approach, and estimated the error due to finite sampling by assuming that we used the trapezoidal rule method to compute the integral. Let $\delta_{i,j,k}$ be the estimate of the smoothed density contrast at the site $(i,j,k)$. The estimator of $\langle \delta^Q \rangle$ is then

$$E(\langle \delta^Q \rangle) = \frac{1}{N_{\rm s}} \sum_{i,j,k} \delta^Q_{i,j,k}. \tag{C6}$$

We estimate the error on this quantity by

$$\Delta_{FS}\langle \delta^Q \rangle \sim \frac{1}{36 N_{\rm s}} \sum_{i,j,k} \left[ |F_{1,i,j,k}| + |F_{2,i,j,k}| + |F_{3,i,j,k}| \right], \tag{C7}$$

with

$$\begin{aligned} F_{1,i,j,k} &= \delta^Q_{i+1,j,k} + \delta^Q_{i-1,j,k} - 2\delta^Q_{i,j,k}, \\ F_{2,i,j,k} &= \delta^Q_{i,j+1,k} + \delta^Q_{i,j-1,k} - 2\delta^Q_{i,j,k}, \\ F_{3,i,j,k} &= \delta^Q_{i,j,k+1} + \delta^Q_{i,j,k-1} - 2\delta^Q_{i,j,k}. \end{aligned} \tag{C8}$$

In the case of a top–hat smoothing, the error was estimated in a more standard way, using the count-in-cell formalism. If we forget discreteness effects to simplify the notations (but we have fully corrected for them in the calculation of $\sigma^2$ and $S_3$), we can write

$$[\Delta_{FS}\langle \delta^Q \rangle]^2 \sim \frac{1}{N_{\rm s}} \left[ <\delta^{2Q}> - <\delta^Q>^2 \right]. \tag{C9}$$

This equation actually assumes implicitly that the cells were thrown at random in the sample, which is not the case in our algorithm. In other words, we certainly overestimate here the error, which should be proportional to $N_{\rm s}^{-1}$ rather than $N_{\rm s}^{-1/2}$.

To estimate the error on $S_3 = \langle \delta^3 \rangle / \langle \delta^2 \rangle^2$, we use the propagation of errors formula, i.e.

$$[\Delta_{\rm XX} S_3]^2 = \frac{[\Delta_{\rm XX}\langle \delta^3 \rangle]^2}{\langle \delta^2 \rangle^4} + 4\frac{[\langle \delta^3 \rangle \Delta_{\rm XX}\langle \delta^2 \rangle]^2}{\langle \delta^2 \rangle^6}, \tag{C10}$$

with XX=FV or FS. After that, we finally write

$$\begin{aligned} {[\Delta S_3]^2} &= [\Delta_{\rm SS} S_3]^2 \tag{C11} \\ &+ \frac{1}{N_{\rm SS}} \sum_{{\rm snapshot}\ S} \left\{ [\Delta_{\rm FV} S_{3,S}]^2 + [\Delta_{\rm FS} S_{3,S}]^2 \right\}, \tag{C12} \end{aligned}$$

where $[\Delta_{\rm XX} S_{3,S}]^2$ stands for the estimate of the error of kind XX for the snapshot $S$.